\documentstyle[epsf,rotate,psfig]{mn}

\def\ang{\mbox{\AA}}

\def\kms{\mbox{kms$^{-1}$}}

\def\deg{\hbox{$^\circ$}}
\def\sig0{$\sigma_0$}

\def\fe52{Fe$_{5270}$}
\def\f53e{Fe$_{5335}$}
\def\ff54e{Fe$_{5406}$}
\def\fg50{Fe$_{5015}$}
\def\m2g{Mg$_2$}
\def\mg2{Mg$_2$}
\def\mg1{Mg$_1$}

\def\asec{\mbox{arcsec}}

\def\ale{\mathrel{\hbox{\rlap{\hbox{\lower4pt\hbox{$\sim$}}}\hbox{$<$}}}}
\def\age{\mathrel{\hbox{\rlap{\hbox{\lower4pt\hbox{$\sim$}}}\hbox{$>$}}}}
\def\etal   {et~al.\ }

\newcommand{\h}[1]{\mbox{$h_#1$}}
\newcommand{\hbeta}[1]{\mbox{H${\beta}$}}

\title[Kinematics, Abundances, and Origin of BCGs]
{Kinematics, Abundances, and Origin of Brightest Cluster Galaxies}

\author[D. Carter et al.]
{D.~Carter$^{1}$, T.J. Bridges$^{2,3,5}$,
\& G.K.T. Hau$^{3,4}$\\
$^1$Liverpool John Moores University, Astrophysics Research Institute, 
Twelve Quays House, Egerton Wharf, Birkenhead, Wirral, L41 1LD.\\
$^2$Royal Greenwich Observatory, Madingley Road, Cambridge, CB3 0EZ.\\
$^3$Institute of Astronomy, Madingley Road, Cambridge, CB3 0HA, UK.\\
$^4$Departamento de Astronomia y Astrofisica, Pontificia Universidad 
Catolica de Chile, Casilla 104, Santiago 22, Chile.\\
$^5$Anglo-Australian Observatory, P.O. Box 296, Epping, NSW 2121, Australia.\\
}

\date{Accepted~~~~~~~.  Received~~~~~~~~~}

\pubyear{1998}

\begin{document}

\maketitle
\begin{abstract}

We present kinematic parameters and absorption line strengths for
three brightest cluster galaxies, NGC 6166, NGC 6173 and NGC 6086.  We
find that NGC 6166 has a velocity dispersion profile which rises
beyond 20 arcsec from the nucleus, with a halo velocity dispersion in
excess of 400 km s$^{-1}$.  All three galaxies show a positive and
constant $h_4$ Hermite moment.
The rising velocity dispersion profile in NGC 6166 thus indicates an
increasing mass-to-light ratio. Rotation is low in all three galaxies,
and NGC 6173 and NGC 6086 show possible kinematically decoupled
cores. All three galaxies have \m2g gradients similar to those found
in normal bright ellipticals, which are not steep enough to support
simple dissipative collapse models, but these could be accompanied by
dissipationless mergers which
would tend to dilute the abundance gradients. The [Mg/Fe] ratios in
NGC 6166 and NGC 6086 are higher than that in NGC 6173, and if NGC
6173 is typical of normal bright ellipticals, this suggests that cDs
cannot form from late mergers of normal galaxies.

\end{abstract}


\section{Introduction}

Brightest cluster galaxies (BCGs) are luminous ellipticals found near
the centres of galaxy clusters and groups.  They are further
subdivided into giant ellipticals (gE), D galaxies which have
shallower surface brightness profiles than gEs, and finally cD
galaxies which have extended low surface brightness envelopes
(Mathews, Morgan \& Schmidt 1964; Schombert 1987, 1988).  BCGs are
often associated with powerful radio sources (Mathews {\sl et al.}
1964), and extended cluster X-ray emission (Jones {\sl et al.} 1979).
In regular clusters, BCGs are invariably at the centre of the cluster
potential, as defined by the cluster X-ray emission, galaxy
distributions, and strong/weak gravitational lensing
(e.g. Miralda-Escude 1995), and their velocities are usually close to
the mean cluster velocity.  BCGs are often observed to have extremely
high globular cluster specific frequencies (e.g. Harris, Pritchet, \&
McClure 1995; Bridges {\sl et al.} 1996).

The origin of BCG halos is one of the most puzzling aspects of galaxy
formation, and it is unclear whether they are formed at the same time
as the galaxy proper, as proposed by Merritt (1984), or are accreted
later.  If accreted later they might form via dissipational processes,
such as star formation in cooling flows (Fabian \& Nulsen 1977), or by
purely stellar dynamical processes, such as galactic cannibalism
(Hausman \& Ostriker 1978; McGlynn \& Ostriker 1980), accretion of
tidally stripped material (Richstone 1976) or ``Galaxy Harassment''
(Moore {\sl et al.} 1996).  For a review of BCG
formation, see Tremaine (1990); see also Garijo, Athanassoula, \&
Garcia-Gomez (1997), and Dubinski (1998).

The kinematics and metallicities of BCG halos provide important clues
as to how they form.  If they form in a dissipational process, then we
might expect a low velocity dispersion, as the energy of the material
from which the stars form has been radiated away.  If BCGs form in a
dissipative collapse, we would expect to see metallicity gradients in
the halo (though metallicity gradients are not unique to dissipative
formation models).  If the halos form out of material from other
cluster galaxies via dissipationless processes, then we might expect
the halos to reflect the kinematics of the cluster as a whole, with
high velocity dispersion and little net rotation; metallicity
gradients would be shallower if BCGs experience significantly more
mergers than other cluster galaxies, since mergers are expected to
weaken metallicity gradients.  If the halos are primordial, and formed
in a process involving violent relaxation, then the kinematics will
reflect the underlying mass distribution.

While there is considerable information on the central kinematics of
ellipticals, there have been few studies of the halo kinematics at
very large radii from optical spectroscopy.  Most such studies have
concentrated on normal ellipticals, and have found that the velocity
dispersion profiles are either flat or decreasing with radius
(e.g. Saglia {\sl et al.} 1993; Carollo {\sl et al.} 1995; Statler,
Smecker-Hane, \& Cecil 1996).  Similar results have been found in the
even fewer studies of BCGs (e.g. Carter {\sl et al.} 1981, 1985; Tonry
1984, 1985; Heckman {\sl et al.} 1985; Fisher, Illingworth, \& Franx
1995).  To date, one cD (IC 1101 in Abell 2029: Dressler 1979; Fisher
{\sl et al.} 1995; Sembach \& Tonry 1996), and two dumbell galaxies
(Abell 3266: Carter {\sl et al.} 1985, and IC 2082: Carter {\sl et
al.} 1981), have been found to have {\it rising} dispersion profiles.

An alternative approach to the study of the kinematics of the outer
halos of cD and other galaxies is the use of globular clusters and
planetary nebulae as tracers. Because of the geometric constraints of
current multi-slit and fibre fed spectrographs, this technique can be
used only in the outer halos, and because of magnitude limits cannot
be applied at distances beyond two or three times the distance of the
Virgo cluster. Nevertheless for M87 (Mould {\sl et al.} 1990; Cohen \&
Ryzhov 1997) and NGC 1399 (Grillmair {\sl et al.} 1994; Kissler-Patig
{\sl et al.} 1998; Minniti {\sl et al.} 1998), globular cluster
velocities have shown conclusively that the M/L ratio increases
outwards in the halo, and thus that there are substantial dark matter
halos in both galaxies.  Arnaboldi {\sl et al.} (1994) confirm this
result for NGC 1399 for a sample of planetary nebulae.  Interestingly,
Cohen \& Ryzhov (1997) find that the globular cluster velocity
dispersion in M87 rises outwards, and in NGC 1399 the velocity
dispersions of the planetary nebulae and globular clusters match
smoothly onto the velocity dispersion of the Fornax cluster itself.
These BCGs at least seem to merge smoothly into their host clusters,
and the BCG and cluster dynamics seem closely linked (see discussion
in Freeman 1997).
The evidence from the distribution of lensed arcs in more distant clusters
points to the dark matter being more centrally concentrated than the X-ray
gas (Miralda-Escude \& Babul 1995). In this case the dark matter will
affect the dynamics of the cD halo significantly.

We have embarked on a program to obtain very deep optical spectroscopy
of BCGs.  Here we present long-slit spectra for NGC 6166, NGC 6173,
and NGC 6086, obtained at the 2.5m INT in La Palma in May 1996.  The
properties of these galaxies and their host clusters are listed in
Table~\ref{tab1}.  NGC 6166 is a luminous cD and a classic
multiple-nucleus galaxy (though the `nuclei' are probably not bound to
NGC 6166), centrally located in a rich cluster with a large cooling
flow ($\sim$ 160 M$_\odot$/yr:~ Allen \& Fabian 1997), and a regular,
symmetrical X-ray appearance (Buote \& Canizares 1996).  Abell 2197 is
1.3 degrees north of Abell 2199 (Gregory \& Thompson 1984), and is a
sparser cluster with an irregular X-ray morphology dominated by two
main concentrations (Muriel, Bohringer, \& Voges 1996).  NGC 6173 is a
D galaxy found at the centre of one of these concentrations
(subclusters), and appears to have a significant peculiar velocity
(i.e. relative to the cluster mean velocity).  Abell 2162 is probably
best described as a poor cluster or compact group; we have not been
able to find any X-ray data for this cluster.  However, NGC 6086 is
classified as a cD galaxy based on its surface brightness profile, and
does not have a peculiar velocity.  Thus, our three galaxies/clusters
span a wide range of properties.

\setcounter{table}{0}
\begin{table*}
\label{tab1}
\caption{Properties of Target Galaxies and Host Clusters.  We list
data for our sample of three BCGs, and two others with rising velocity
dispersions, IC 1101 in Abell 2029 and the giant dumbell in Sersic
40/6 (Abell 3266).  Column 1 designates the galaxy and its host
cluster.  Column 2 gives the galaxy morphological type from Schombert
(1986), except for Sersic 40-6.  Column 3 gives the total absolute
blue magnitude assuming H$_0$=75 km s$^{-1}$Mpc$^{-1}$, and taking B$_T$ values
from Burstein {\sl et al.} (1987), and Peletier {\sl et al.} (1990)
for IC 1101.  Column 4 gives the central velocity dispersion: Burstein
{\sl et al.} (1987) for N6166, N6173, and N6086; Carter {\sl et al.}
(1985) for Sersic 40-6; Carter {\sl et al.} (1981) for IC 2082;
Fisher {\sl et al.} (1995) for IC 1101.
Columns 5, 6 and 7 give the galaxy velocity, cluster velocity,
and cluster velocity
dispersion, respectively: Zabludoff {\sl et al.} (1993a) for
N6166/A2199 and N6173/A2197; Postman \& Lauer (1995) and Zabludoff
{\sl et al.} (1993b) for N6086/A2162; Oegerle, Hill, \& Fitchett
(1995) for IC 1101/A2029; Teague, Carter, \& Gray (1990) for Sersic
40-6/A3266; Ellis {\sl et al.} (1984) for IC 2082.  
Column 8 gives the Bautz-Morgan classification, and
Column 9 gives the cluster richness.  Column 10 contains information
about the cluster's X-ray appearance: `CF' means a cooling flow is
present; `Reg'/`Irr' means the cluster has a regular/irregular X-ray
morphology.}
\begin{tabular}{cccccccccc}
\hline
Galaxy/Cluster & Galaxy & M$_{B_T}$ & $\sigma_0$ &
V$_{gal}$ & V$_{cluster}$ & $\sigma_{clus}$ & BM & R & X-ray \\
 & Type & & & (km s$^{-1}$) & (km s$^{-1}$) & & & \\
 \\

 & & & Our & Sample & & & & \\ \\

N6166/A2199 & cD & $-$22.7 & 326 & 9293 &9063& 823 & I & 2 & CF, Reg \\
N6173/A2197 & gE & $-$22.4 & 261  & 8800 &9134& 550 & III & 1 & Irr \\
N6086/A2162 & cD & $-$21.8 & 304 & 9547 &9795& 302 & II-III & 0 & ? \\ \\
 
 \\

IC 1101/A2029 & D & $-$23.5  & 359 & 23399 & 23169 & 1436 & I & 2 & CF, Reg \\
Sersic 40-6/A3266 & cD &  & 327 & 17914 & 17802 & 1186 & I-II & 2 & CF, Reg \\
IC2082/AS0463& D & &265&12051&12035&844& I-II & 0 & ? \\
\hline
\end{tabular}
\label{tab1}
\end{table*}

NGC 6166 has previously been observed spectroscopically by Tonry
(1985) and Fisher {\sl et al.} (1995) with 2 and 3 position angles
respectively (each include the major axis).  Unfortunately, both
studies only extend to 10$-$20 arcsec from the galaxy centre, and over
this limited range both find a flat or slightly decreasing velocity
dispersion profile (we show below that this is consistent with our
data, which extend much further out).  Fisher {\sl et al.}  find that
NGC 6166 does not have significant rotation along either the major or
minor axes out to 20 arcsec radius.  We have only been able to find
central velocity dispersions and line strengths for the other two
galaxies (e.g. Burstein {\sl et al.} 1987; McElroy 1995).

\section{Observations and Data Reduction}

Table~\ref{tab2} contains details of the observations.  Spectra are in
each case taken with the slit aligned with the major axis to maximise
the signal to noise ratio. The spectra include the Mg and Fe
absorption lines, as well as [OIII] and [NI] emission lines (for NGC
6166 only).  We also obtained template spectra of 28 stars, with
spectral type ranging from G6 to M1, and luminosity class III, IV and
V.

\begin{figure*}
\label{fig1}
  \centerline{ \hbox{
   \psfig{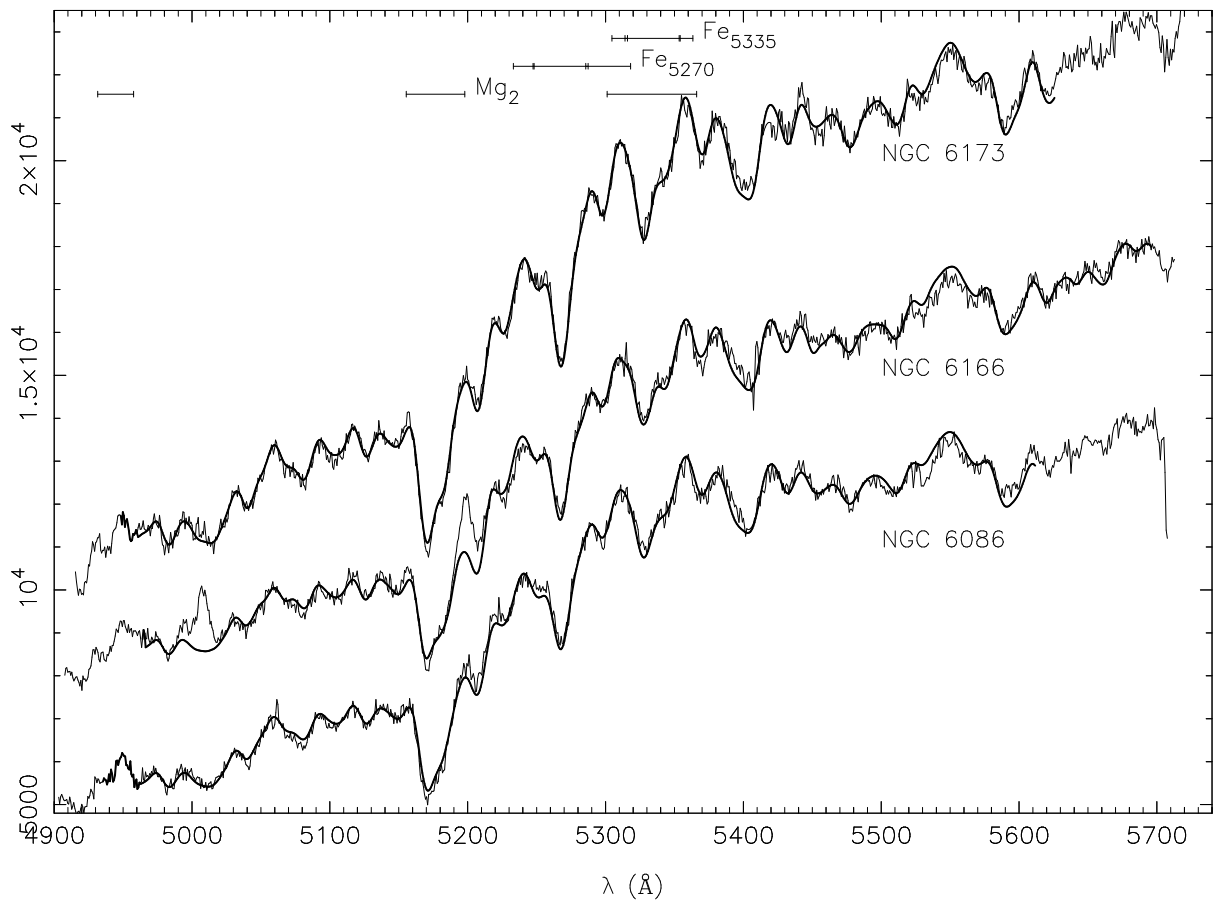}}}
\caption{The central galaxy spectra for NGC 6166, NGC 6086, and NGC
6173, in the rest frame of the galaxies.  The lighter solid lines show
the data, while the heavier solid lines are the best-fitting galaxy
model for each galaxy.  The scale is arbitrary.  We have also
indicated the definitions of the Lick \fe52 and \f53e bandpasses, and
our redefined \m2g index.}
\label{fig:censpec}
\end{figure*}

Figure~\ref{fig:censpec} shows the central spectra for each of the
three galaxies; for each spectrum, the best-fitting stellar template
is also shown (see below), as well as the bandpass definitions for
\m2g, \fe52, and \f53e.  Preliminary data reduction was done using
standard FIGARO routines, and followed the procedures of Carter,
Thomson and Hau (1998). This involves debiassing, flatfielding and
wavelength calibrating the data.  A slope in the spectrum due to
detector rotation was corrected using the S-distortion correction
routines in FIGARO.  The template spectra were re-binned to zero
velocity, using their published radial velocities, mainly from Wilson
(1953).

\setcounter{table}{1}
\begin{table}
\label{tab2}
\caption{Observing Log and Instrumental Setup}
\begin{tabular}{cc}
\hline\\
Telescope & Isaac Newton Telescope (2.5 m) \\
Instrument & Intermediate Dispersion Spectrograph \\
Observing dates & 10--16/06/1996 \\
Camera & 235mm \\
Grating & R1200Y \\
Dispersion & 35 {\AA}/mm \\
Detector & Tektronix 1024 CCD \\
Resolution & 2 \AA \\
Spectral Coverage & 4930--5730 \AA \\
Slit Width & 3.0 arcsec \\
Average Seeing & 1.3 arcsec \\
Slit PA & 41/140/0 deg$^1$ \\
Exposure Time & 18.75/6.25/6.33 hrs$^1$ \\

\hline

\end{tabular}
\medskip

\noindent$^1$~~For NGC 6166/6173/6086 respectively.

\end{table}

Following van der Marel \& Franx (1993), the line-of-sight velocity
distributions (LOSVDs) are modelled as Gaussians with small Hermite
deviations, parameterised as mean velocity $v$, dispersion $\sigma$,
and hermite moments $h_n$ of order $n$. The parameters are recovered
by the program {\tt kinematics} written by H-W. Rix; for more details
please see van der Marel \& Franx (1993), Rix \& White (1992), and
Hau, Carter, \& Balcells (1998).  The Gauss-Hermite moments \h3 and
\h4 are closely related to the skewness and kurtosis of the LOSVD
respectively, and are therefore easy to visualise.  Large \h3 with
opposite sign to the mean velocity is typical of rapidly-rotating
systems, whilst for spherical two-integral models, \h4 is proportional
to the velocity dispersion anisotropy $\beta$ (van der Marel \& Franx
1993). The higher moments are more difficult to interpret. However,
they are useful for identifying structures in the residuals of the
model fit to the data. We measure and plot Gauss-Hermite parameters up
to \h6.

The observed wavelength range is well known to give template mismatch
problems (van der Marel \& Franx 1993).  In order to minimize template
mismatch, an optimal stellar template is employed for the direct
fitting. This is estimated as the best linear combination of the 28
template spectra which when broadened with a Gaussian profile
minimizes $\chi^2$, the difference between the combination and
the galaxy spectrum, after an initial guess of the velocity for each template
and the dispersion, (Rix \& White 1992).  
This optimal template is then used for recovering the
LOSVD for the galaxy spectrum, modelled up to \h6.  This procedure is
done for each of the individual spectra in each galaxy, thus reducing
radial template mismatch due to abundance gradients.  The galaxy
nucleus is located by a Gaussian fit to the light profile of the
central few pixels.  Monte Carlo experiments show that 2500
counts/pixel or higher
gives satisfactory results for the direct fitting (see 
Appendix A).  

Our procedure allows the determination of an individual optimal composite
template for each data point, To ensure that this procedure does not
introduce systematic effects into our profiles we repeated the analysis 
for NGC 6166 using the composite template for the central spectrum for
all data points. There are no systematic differences in zeropoint or overall
trend in any of the kinematic values derived. Subjectively, we estimate that
the scatter is higher, justifying our decision to use individually determined
optimal templates.

Emission lines could affect the measurement of absorption-line indices
if they fall within one of the bandpasses, and also can potentially
introduce systematics in the absorption line profile analysis.
Significant [OIII] and [NI] emission are observed in the nuclear
region $r \ale 11\ $ \asec of NGC 6166, but not in NGC 6173 or NGC
6086.  Thus, for NGC 6166, pixels within a wavelength difference
corresponding to 3$\sigma$ from an emission line are excluded from the
line-profile fitting.  These regions are: $4988$--$5021\ang$ ([OIII]),
$5179$--$5213\ang$ ([NI]), and $5397$--$5415\ang$ (strong sky
line). Experimentation shows that excluding a few pixels has a
negligible effect on the recovery of the LOSVD as kinematics are
measured over the entire spectrum.  For spectra more than 11 arcsec
from the nucleus where no emission is present, only the region
affected by the strong sky line is excluded, allowing fitting over the
entire spectrum.  Another difficulty with NGC 6166 is the presence of
companion nucleus B, which is at a PA of 62\deg with respect to the
central cD, and contributes significantly to the light in our slit
between $\sim$ 5$-$20 arcsec from the galaxy centre along the NE side.
{\tt Kinematics} was not designed to recover two LOSVDs, so we
restrict our analysis to the SW side of the galaxy.

The parameters recovered by the Gauss-Hermite analysis are presented
in Figures~\ref{fig:hermplot}(a), (b), and (c), for NGC 6166, NGC
6173, and NGC 6086 respectively.

\setcounter{figure}{1}
\begin{figure*}
  \centerline{ \hbox{
   \psfig{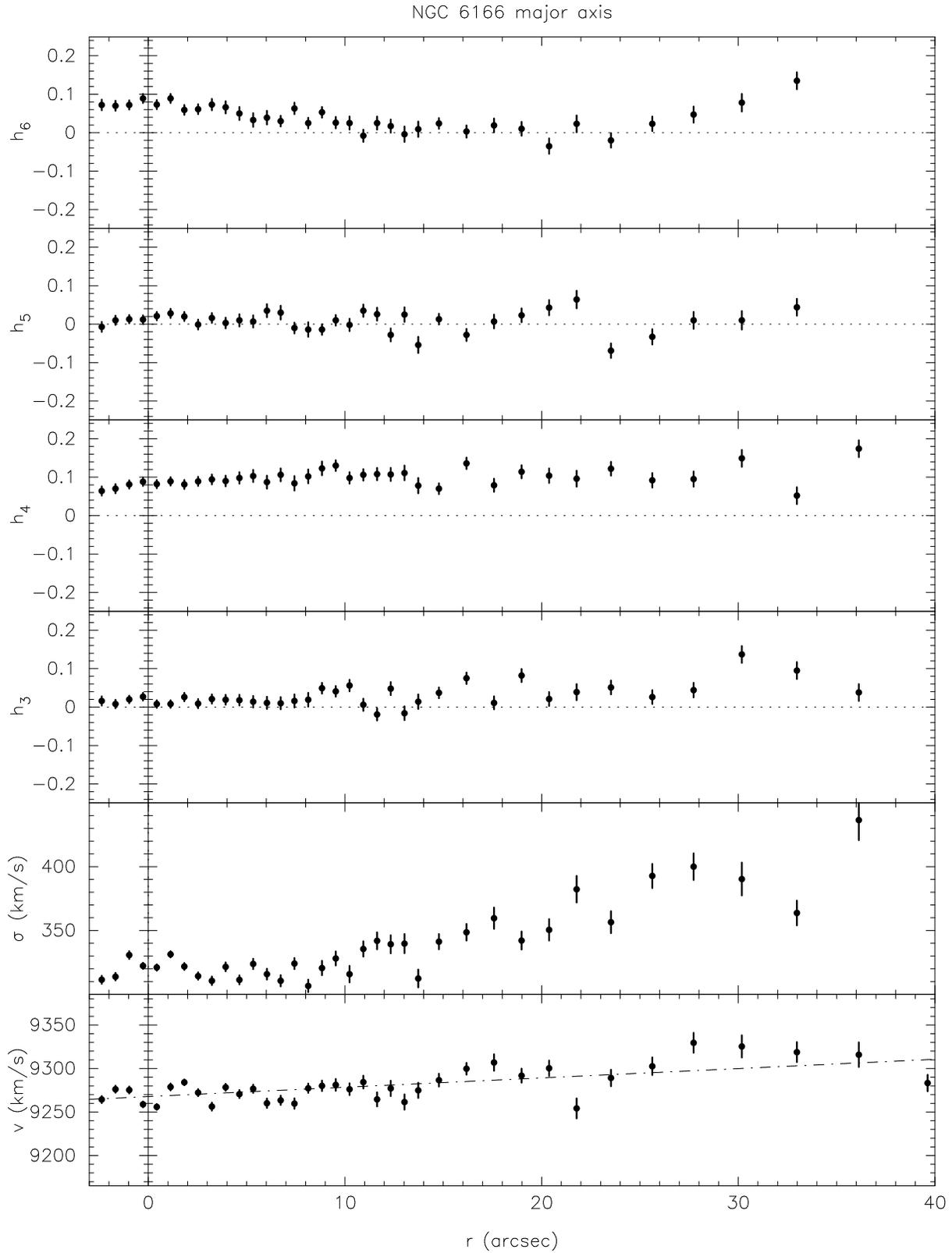}}}
\caption{{\bf (a)}~The $v,\sigma,h_3,h_4,h_5$ \& $h_6$ along the major axes 
for NGC 6166.  The $1 \sigma$ errorbars are also plotted.  A weighted
least-squares fit to the velocity profile is shown, with a slope 
of 1.07 km s$^{-1}$ arcsec$^{-1}$.}  
\label{fig:hermplot}
\end{figure*}

\setcounter{figure}{1}
\begin{figure*}
  \centerline{ \hbox{
   \psfig{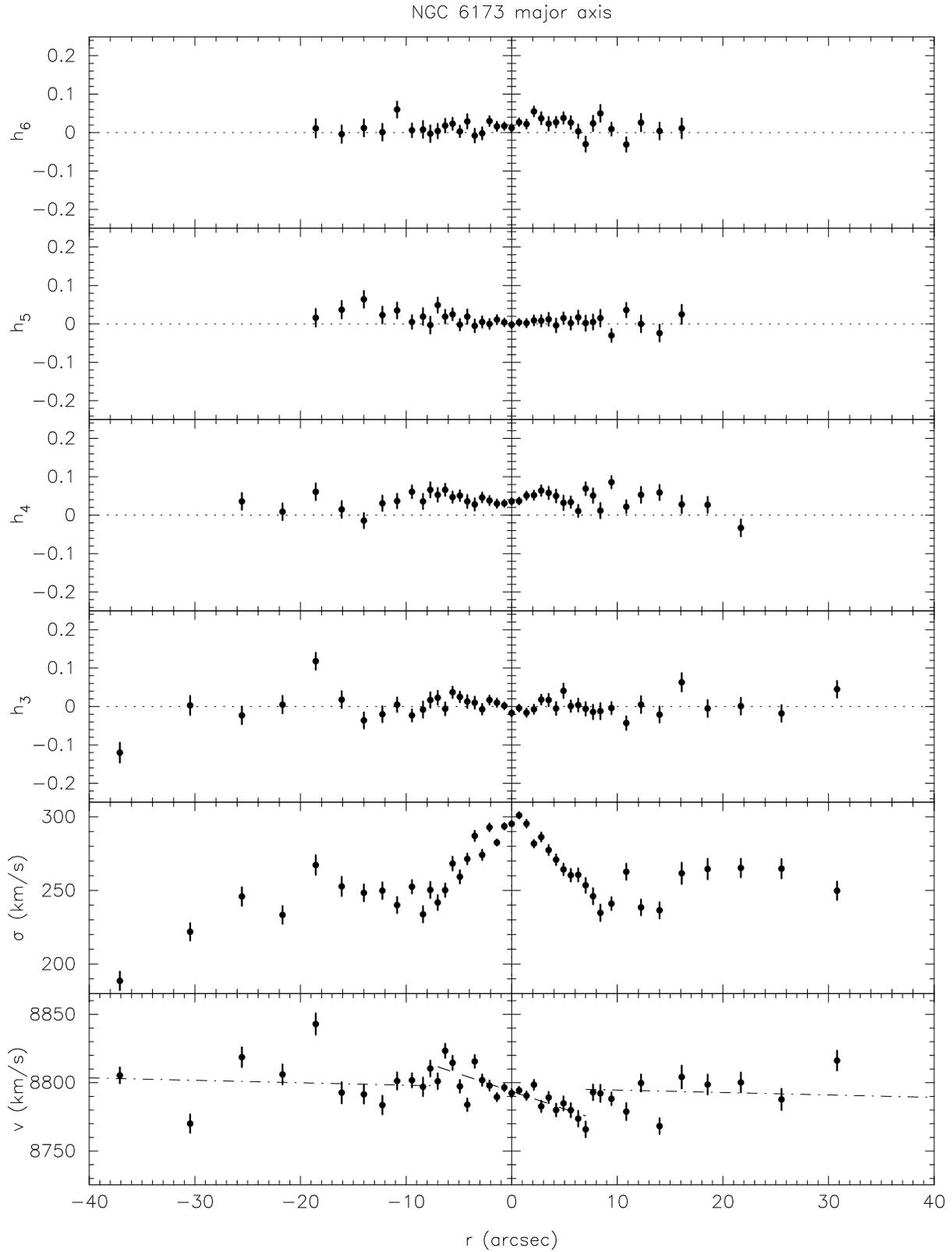}}}
\caption{{\bf (b)}~Same as Figure~\ref{fig:hermplot}(a), except for
NGC 6173.  For $-$40 $<$ r $\leq$ $-$7 and 7 $<$ r $<$ 40 arcsec,
the velocity gradient is $-$0.18 km s$^{-1}$ arcsec$^{-1}$.  For 
7 $<$ r $\leq$ 7 arcsec, the gradient is $-$2.56 km s$^{-1}$ arcsec$^{-1}$.}
\label{fig:hermplot}
\end{figure*}

\setcounter{figure}{1}
\begin{figure*}
  \centerline{ \hbox{
   \psfig{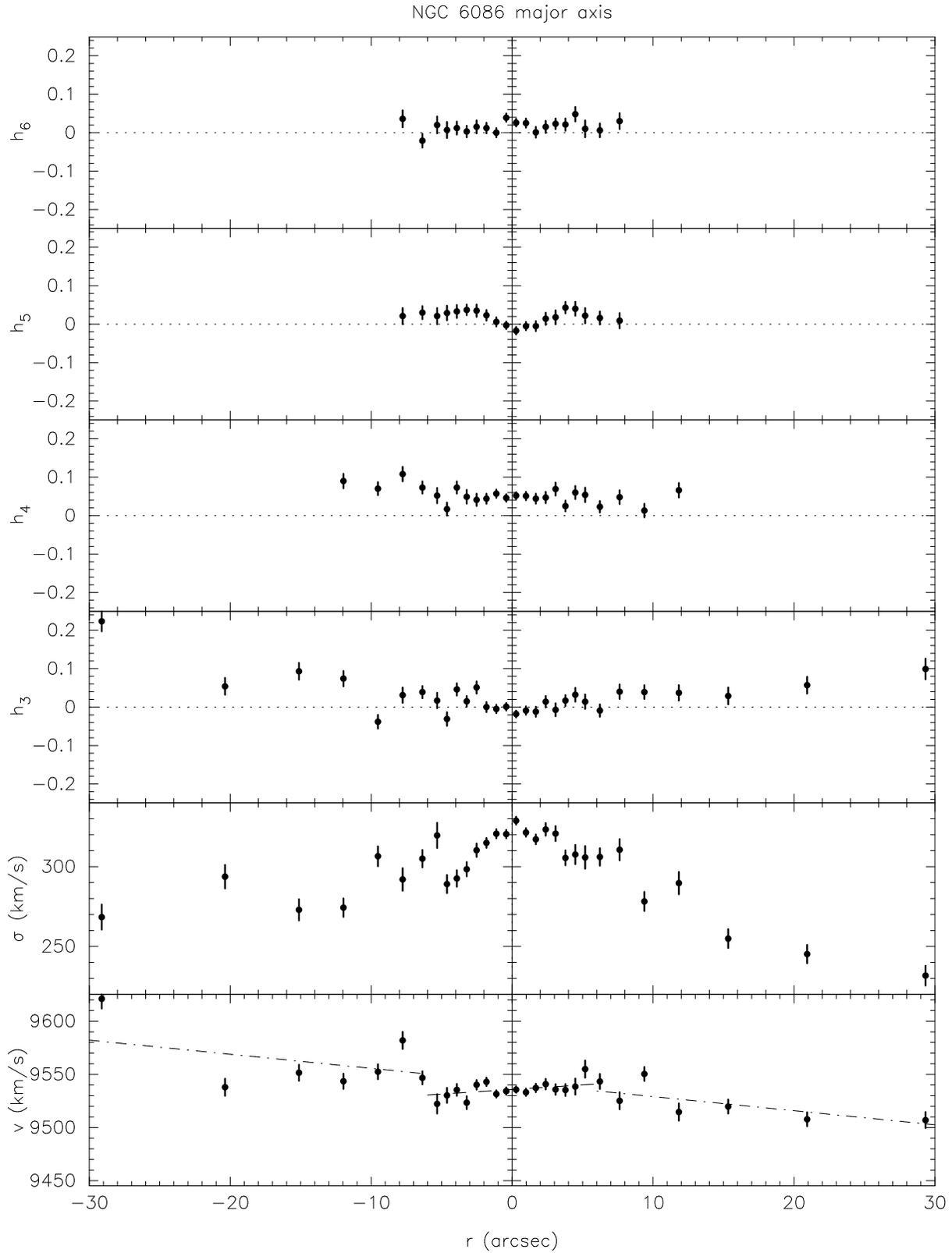}}}
\caption{{\bf (c)}~Same as Figure~\ref{fig:hermplot}(a), except for
NGC 6086.  For $-$6 $<$ r $\leq$ 6 arcsec, the velocity gradient is
0.86 km s$^{-1}$ arcsec$^{-1}$, and for the points outside 6 arcsec, the gradient
is $-$1.33 km s$^{-1}$ arcsec$^{-1}$.}
\label{fig:hermplot}
\end{figure*}

\section{Stellar Kinematics}

\subsection{Velocity Dispersions}

\noindent{\it Central Velocity Dispersions}

\medskip

We can compare our central velocity dispersions, $\sigma_0$ = 320
$\pm$ 5, 295 $\pm$ 5, and 325 $\pm$ 8 km s$^{-1}$ for NGC 6166, 6173, and
6086 respectively, with previous determinations.  NGC 6166 has been
the best-studied of the three:~ Malumuth \& Kirshner (1981), Tonry
(1985), Burstein {\sl et al.} (1987), and Fisher {\sl et al.} (1995),
find $\sigma_0$= 350 $\pm$ 35, 303, 326, and 295 $\pm$ 6 km s$^{-1}$
respectively.  For NGC 6173, Burstein {\sl et al.} (1987) and McElroy
(1995) find $\sigma_0$= 261 and 240 km s$^{-1}$ respectively, both values
lower than ours.  For NGC 6086, Burstein {\sl et al.} (1987) and
McElroy (1995) find $\sigma_0$= 304 and 325 km s$^{-1}$
respectively. Overall, the agreement is good, and the present data are
of much higher quality than any previous data.

\medskip

\noindent{\it Velocity Dispersion Profiles}

\medskip

From Figures~\ref{fig:hermplot}(b) and (c), we see that both NGC 6173
and NGC 6086 have velocity dispersion profiles that decline outwards.
Such behavior is found in most normal ellipticals.  In contrast,
Figure~\ref{fig:hermplot}(a) shows that the velocity dispersion
profile of NGC 6166 {\it increases} from $\sim$ 325 km s$^{-1}$ at the
centre to $\sim$ 450 km s$^{-1}$ at 35 arcsec along the major axis.  Tonry
(1985) and Fisher {\sl et al.} (1995) both have limited dispersion
profiles for NGC 6166, and find a flat and slightly decreasing
dispersion profile respectively, out to $\sim$ 20 arcsec along the
major axis.  This is consistent with what we find, since our
dispersion only starts rising significantly beyond $\sim$ 20 arcsec.

In Appendix A we analyse possible systematic errors introduced into the
kinematic parameters by noise and imperfect sky subtraction. 
We find that there is a bias towards low velocity dispersion when the
signal-to-noise is poor, but this does not significantly affect the points
plotted in Figure 2(a-c). 

\subsection{Rotation}

From Figure~\ref{fig:hermplot}(a), NGC 6166 shows modest major-axis
rotation, and a weighted least-squares fit yields a slope of 1.07
$\pm$ 0.12 km s$^{-1}$ arcsec$^{-1}$, or a rotation of $\sim$ 45 km s$^{-1}$ at 40
arcsec.  Tonry (1985) found a rotation amounting to 4 km s$^{-1}$ arcsec$^{-1}$ at
PAs of 62\deg and 102\deg, while Fisher et al (1995) find no
significant rotation at PAs of 31\deg, 35\deg and 125\deg.

NGC 6173 and NGC 6086 both show evidence of rotation in the core,
decoupled from the kinematic properties further out. In NGC 6173
(Figure~\ref{fig:hermplot}(b)), a linear least-squares fit gives a
velocity gradient of $-$2.56 $\pm$ 0.27 km s$^{-1}$ arcsec$^{-1}$ for
$|r| < 7$ arcsec, with no significant rotation outside this, and in
NGC 6086 (Figure~\ref{fig:hermplot}(c)), similar fits give a velocity
gradient of 0.86 $\pm$ 0.48 km s$^{-1}$ arcsec$^{-1}$ for $|r| < 6$
arcsec and $-$1.33 $\pm$ 0.13 km s$^{-1}$ arcsec$^{-1}$ for $6 < |r| <
32$ arcsec.  Although the core rotation is only marginally
significant, it does appear to be counter-rotating.  NGC 6173 is a
shell galaxy, and both the shells and higher core rotation could be a
result of a merger or interaction.  However, in both galaxies, there
is no significant asymmetry in the line profiles (i.e. changes in
$h_3$) associated with the peculiar kinematics, as expected from
classical KDCs like IC 1459 or NGC 2865 (Hau, Carter, \& Balcells
1998).  Also, Statler (1991) proposes that KDCs might also arise from
streaming in a triaxial potential, without a merger/interaction.

\subsection{$h_4$}

The parameter $h_4$ measures symmetric deviations from a Gaussian
profile and is related to the velocity anisotropy. $h_4$ is
significantly positive, with no radial dependence, in all of our
galaxies. Although it is not straightforward to translate $h_4$ to the
velocity anisotropy parameter $\beta$ ($=1 -
\sigma_\theta^2/\sigma_r^2$), positive values, at least outside
the nucleus, generally indicate a bias towards radial
orbits (Gerhard 1993; Gerhard {\sl et al.} 1998, Rix {\sl et al.}
1997). The most important result from the line-profile analysis is
that there is no change in $h_4$ associated with the rising dispersion
in NGC~6166, indicating that the increase in velocity dispersion is
not associated with a change in velocity anisotropy towards tangential
orbits.

$h_4$ has a magitude of 0.10, 0.05 \& 0.05 in NGC~6166, 6173 \&
6086 respectively. Template mismatch can introduce systematics in
$h_4$ by either increasing or decreasing its magnitude; see, for
example, van der Marel {\sl et al.} (1994), and Hau, Carter, \&
Balcells (1998).  The $\chi^2$ fits in NGC~6173 \& NGC~6086 appear
to be near perfect.  Inspection of the residuals for NGC 6166 shows
that the fits are less perfect than for the other two galaxies, but
are satisfactory.  We feel that template mismatch is not a major
problem for all three galaxies as the optimal template is estimated
from 28 template stars.  

In order to estimate the effect of template mismatch on the absolute value
of $h_4$ (and other kinematic parameters), in Appendix A we derive the 
kinematic parameters for NGC 6166 for a range of individual templates.
None of these templates fit the galaxy data as well as the optimal template.
Given an extreme range of templates, we find zero point differences of up to 
0.1 in $h_4$, $h_5$ and $h_6$; 0.2 in $h_3$, and 50 km/s in velocity 
dispersion. However these differences are largely in the zero points for
stars at the extreme end of the range of spectral types of the observed 
templates (G6 III and M1 III). In particular low $h_4$ is found only when an 
M1 III template is used, this template does not fit the data well and 
contributes little to the optimal composite templates. Although
template mismatch may in principle shift the zero-point in
$h_4$, it cannot explain the significantly positive $h_4$ seen in our
data.

In Appendix A we also analyse the effect of noise on the recovery of $h_4$
when the magnitude of the real value of $h_4$ is large. We find that there
are no systematic biases unless $|h_4| > 0.1$, when $|h_4|$ is systematically
underestimated at low signal to noise. Thus the only possible bias in our
results would be to underestimate $h_4$ in the outer regions of NGC 6166,
although there is no evidence that this has happened, we cannot absolutely 
rule it out.

Van der Marel \& Franx (1993) give formulae which relate $h_4$ to
$\beta$, but they are model-dependent.  To better quantify the
anisotropy parameter $\beta$, we need to solve for the $M/L$ ratio
using all of the available kinematic and photometric data. Such
techniques have been developed by Rix {\sl et al.} (1997), using
Schwarzchild's method of populating orbits, and by Dehnen (1995)
for the axisymmetric case in order to model M32.

In summary, the rising velocity dispersion profile in NGC~6166 is not
due to a change in velocity anisotropy, but rather, reflects an
increasing $M/L$ ratio. This indicates that NGC~6166 possesses a
massive dark halo.

\section{Gas kinematics in NGC 6166}

\begin{figure*}
  \centerline{
  \hbox{\psfig{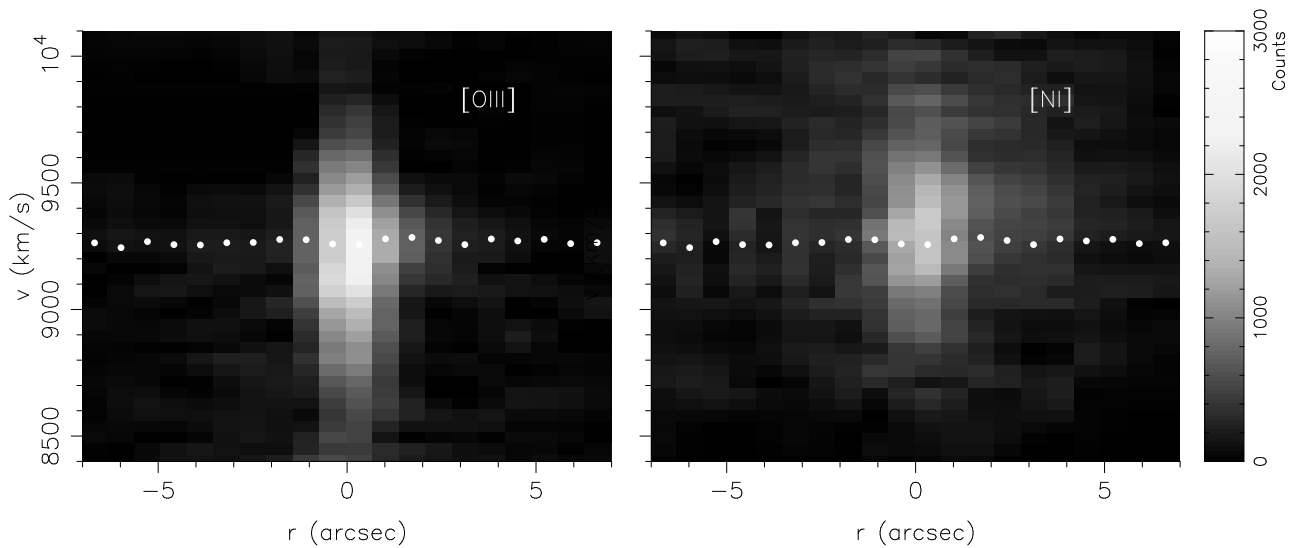}} }
\caption{The gas kinematics from the [OIII] and [NI] emission lines at
$5007\ang$ and $5200\ang$ along the major axis of NGC 6166 are ploted
in the left and right plots respectively. The stellar velocity curves
from Fig.~\ref{fig:hermplot}(a) are over-plotted in white. }
\label{fig:gaskin}
\end{figure*}

Significant [OIII] and [NI] emission is observed in the centre of
NGC~6166.  In this section we present the line-of-sight velocity
distribution of the ionized gas inferred by the profile of the [OIII]
and [NI] emission lines, obtained directly from the residual map of
the Gauss-Hermite analysis, described in the previous section.  This
method is very successful and is more objective than fitting the
baseline by hand. The results are plotted in Figure~\ref{fig:gaskin},
with the stellar rotation curve from Figure~\ref{fig:hermplot}(a)
over-plotted.  We see that the [OIII] and [NI] emission is very
concentrated to the galaxy centre, with spatial FWHM of 1.9 and 3.0
arcsec respectively. The spatial FWHM of template stars is about 2.3
arcsec, thus the [OIII] emission region is unresolved but that of the
[NI] may be marginally resolved.  At the galaxy centre, the [OIII] and
[NI] have velocity dispersions $\sigma$ of $333 \pm 11$ and $365 \pm
20$ km s$^{-1}$ respectively, which is very similar to the central stellar
velocity dispersion (Figure 2a).

\section{Stellar Absorption Line Abundances}

\subsection{Introduction}

In this section the absorption line indices are presented for the
three galaxies.  Mg$_2$, \fe52, and \f53e are extracted from
individual, de-redshifted spectra following the definitions and
recipes of Faber \etal (1985), and adopting the passband definitions
of Worthey \etal (1994); see Appendix B for more details.  We first
discuss the local Mg$_2-<$Fe$>$ relations for each galaxy, then present
the Mg$_2$ radial profiles.  We discuss the significance of these
results in our Conclusions.

As discussed in Appendix B, there is some uncertainty regarding the
zero-point of our measured indices in the Lick/IDS scale, due to the
fact that only one star with Lick indices (HR 5227) has been observed,
and also that the grating settings for the galaxy and the template
star observations were different. However recent observations by
Trager {\sl et al.} (1998) for the nucleii of NGC 6166 and NGC 6086
do agree within their admittedly rather large error bars.
We also find 
that our \m2g measurements agree well with those measured by Cardiel
{\sl et al.} (1997), if both are not zero-point corrected (Figure
~\ref{fig:uscardiel}).
As most of our discussion is not based on the absolute values of the
indices, throughout this Section we present line indices without
zero-point corrections.

\subsection{Local $<$Fe$>$ Versus Mg$_2$}

\begin{figure*}
   \centerline{ \hbox{
   \psfig{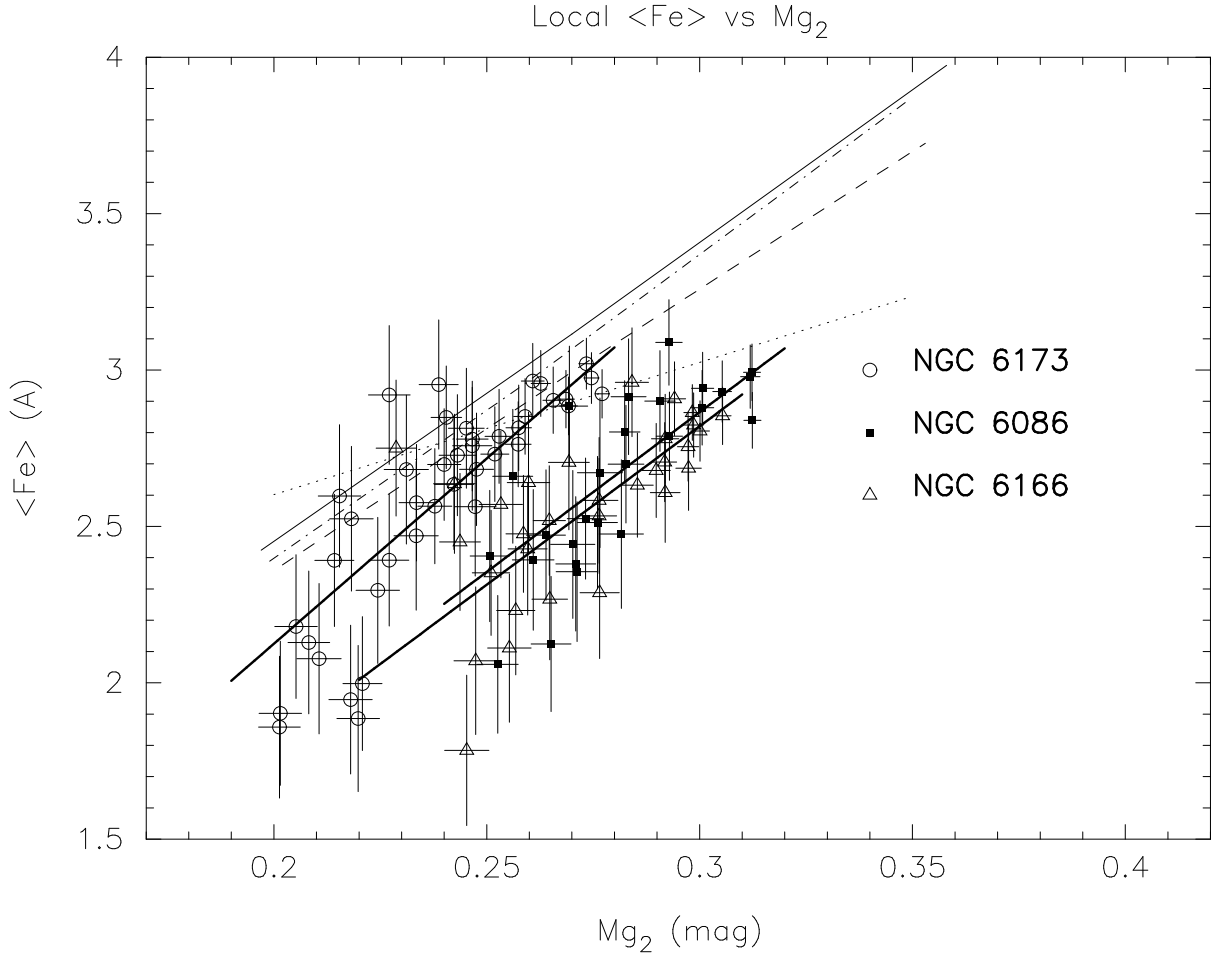}}}
\caption{Local $<$Fe$>$ Versus Mg$_2$ for three BCGs.  Open triangles
are plotted for NGC 6166, filled squares for NGC 6086, and open
circles for NGC 6173.  No zeropoint corrections have been applied to
the data in Figures 4, 5 and 6.  For NGC 6173 and NGC 6086, the data
have been folded about the major axis, while for NGC 6166 we only show
the SW side of the major axis, to avoid contamination from the
secondary nucleus on the NE side.  The dark solid lines are weighted
linear least-squares fits to the data points; from left to right, they
correspond to NGC 6173, NGC 6086 and NGC 6166.  Plotted also are the
locations expected for a 12 Gyr old population with 0.4 $<$
Z/Z$_\odot$ $<$ 2.5, predicted by 3 population synthesis models:
Worthey (1994; thin solid line); Bruzual \& Charlot (1997; dot-dashed
line), and Vazdekis {\sl et al.} (1996; dashed line).  The line fitted
to the nuclear indices of a sample of galaxies from Worthey, Faber, \&
Gonzalez (1992) is indicated by the dotted line.}
\label{fig:fevsmg2}
\end{figure*}

In Figure~\ref{fig:fevsmg2}, we plot $<$Fe$>$ [$\equiv$ (\fe52 +
\f53e)/2)] against \m2g for the three galaxies.  All three galaxies
appear to have [Mg/Fe] ratios higher than the model values with solar
[Mg/Fe] abundance, but there are uncertainties in our zero-points.
Notice further from Figure~\ref{fig:fevsmg2} that NGC 6086 and NGC
6166 are more `enhanced' than NGC 6173.  This is interesting, given
that the first two galaxies are centrally-located cDs, while NGC 6173
is best described as a gE in a subcluster in Abell 2197.  Similarly,
M87 is offset from the non-BCGs in the data of Davies, Sadler, \&
Peletier.

Hau (1998) has pointed out that in galaxies of high velocity
dispersion, a positive value of $h_4$, such as might result from
radial anisotropy, leads to more absorption being lost from the
passband that the Fe indices are measured from, and thus to an
artificially high value of [Mg/Fe]. This systematic error might
account for as much as a third of the offset between NGC 6166 and NGC
6173. However we still find that our galaxies with cD morphology have
higher [Mg/Fe] compared with NGC 6173, which is the only normal bright
elliptical in our sample.  If further observations confirm that NGC
6173 is typical (despite uncertainty in the relative zero-points, the
data of Davies {\sl et al.} (1993) support this), then there is a
difference between cDs and normal bright ellipticals.  Bender (1996)
uses the offset of [Mg/Fe] to higher values in ellipticals when
compared with spirals as an argument against the idea that ellipticals
form from the mergers of objects similar to {\it present-day} spirals;
any such argument is clearly stronger for cDs.  Bender (1996) and
Davies (1996) feel that the most plausible explanation for the
non-solar [Mg/Fe] ratios in ellipticals is a short burst of star
formation, $<$ 1$-$2 Gyr.  This would produce a population enriched in
Mg from rapid Type II supernovae from massive stars, while few stars
would be formed from Fe-enriched gas formed in longer-lived Type Ia
supernovae.  See Section 6 for further discussion on these topics.

\subsection{Mg$_2$ Abundance Gradients}

\begin{figure*}
  \centerline{ \hbox{
   \psfig{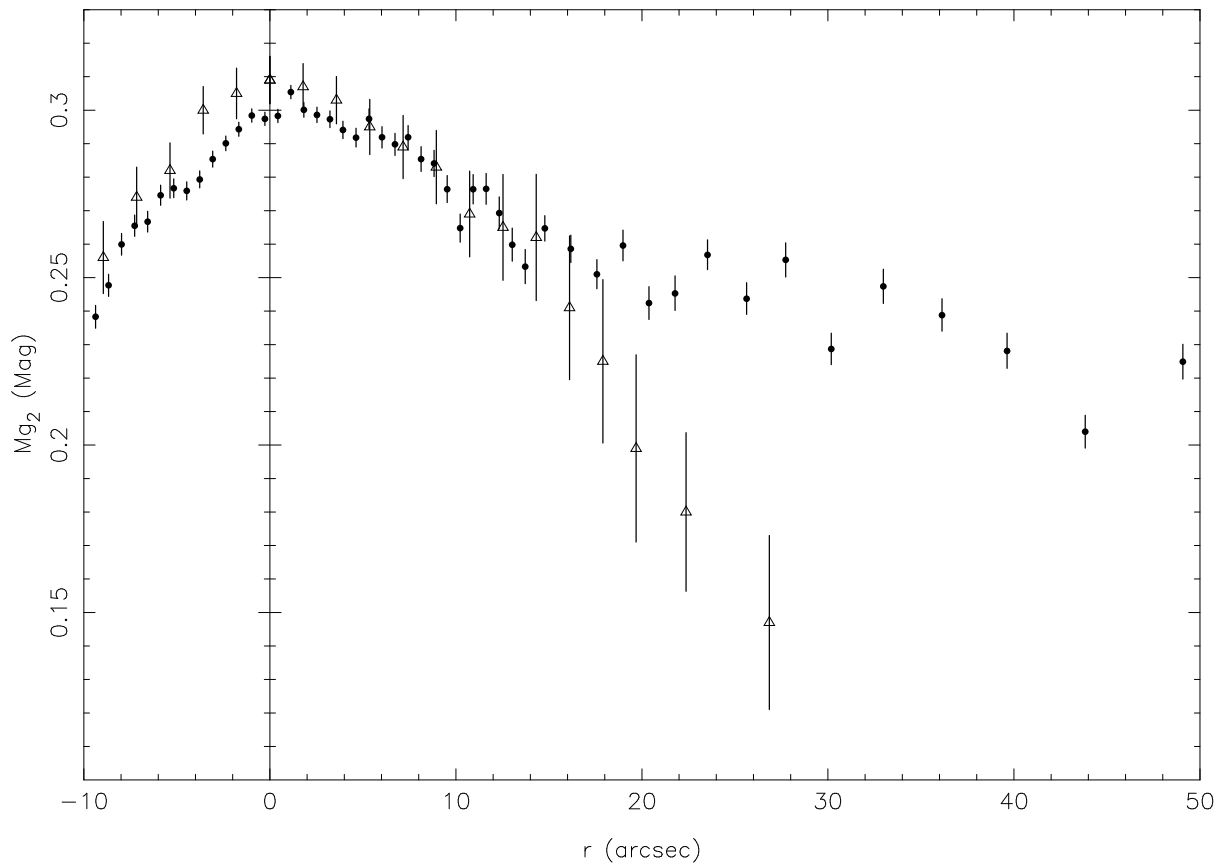}}}
\caption{Mg$_2$ profiles for NGC 6166.  \m2g is plotted against radius
along the SW side of the major axis.  Closed circles are plotted for
our data, while open triangles are used for the data of Cardiel
{\sl et al.} (1997).  Both datasets have not been zero-point corrected.} 
\label{fig:uscardiel}
\end{figure*}

We first compare our Mg$_2$ profile for NGC 6166 with that of Cardiel
{\sl et al.} (1997) in Figure ~\ref{fig:uscardiel}.  We have only
plotted the SW side of the major axis because of contamination from
the secondary nucleus on the NE side.  Neither set of data has had any
zeropoint corrections applied (see Appendix B).  Out to $\sim$ 15
arcsec, the agreement is very good.  Beyond 15 arcsec, our \m2g is
significantly higher than that of Cardiel {\sl et al.}  The source of
this discrepancy is unclear, however 
our signal-to-noise is higher than that of Cardiel {\sl et al.}  Their
three outermost points have \m2g values $<$ 0.2, lower than any of the
ellipticals studied by Davies {\sl et al.} (1993).  These extreme
values seem unlikely, and we are confident of the reliability of our
\m2g profiles.  

The sensitivity of our line strength measurements to sky subtraction errors
needs to be considered. The sky spectrum that we subtract is averaged over
many spatial pixels and is relatively low noise. The only feature in the sky
spectrum are the night sky emission line at 5577{\AA} and (much weaker)
5199{\AA}, apart from this the sky is well represented by a constant 
continuum. At these redshifts the [NI] line at 5199{\AA} does not fall in
the \m2g band or its continuum bands, and so can be neglected. Therefore
the sky in the region of all of the line and continuum bandpasses can be 
represented by a constant, and the error in this constant can be determined
from the maximum residuals seen in the [OI] line at 5577{\AA} in the sky 
subtracted spectra, and the ratio of the peak intensity in this line to the 
continuum in the sky spectra.

Inspection of the sky line residuals in the outer regions
of our profile show that sky subtraction errors do not exceed 1\% of the 
signal level, thus systematic errors in our \m2g values will not exceed 
0.01 mag.

In Figure~\ref{fig:mg2prof}, we present the major axis profiles of
\m2g, for NGC 6166, 6173, and 6086.  Outside 3--5 arcsec, the \m2g
profiles are well-fit by power-laws of slope (d(\m2g)/d(log $r$))
$-$0.080, $-$0.059, and $-$0.073 for NGC 6166, 6086 and 6173
respectively (inside 3 arcsec, seeing effects are likely to be
important; the flattening of \m2g in NGC 6166 within a larger radius
of $\sim$ 5 arcsec may be due to star formation inside the
emission-line region in this galaxy).  These slopes are fairly similar
to each other, and are also similar to the average \m2g slope of
$-$0.058 found by Cardiel {\sl et al.} (1997) for eight BCGs without
cooling flows or optical emission lines.
Gorgas, Efstathiou, \& Aragon-Salamanca (1990) also measured shallower
gradients for two BCGs ($-$0.051 and $-$0.021 for the cDs in 0559-40
and PKS 2354-35).  However, examination of their Figure 9 shows that
these gradients are based on only three points, and the outermost
point in each case has large uncertainty.

Cardiel {\sl et al.} (1997) find that the \m2g slopes are shallower in
BCGs with optical line-emission, inside the emission-line region; our
data for NGC 6166 are consistent with this.  Cardiel {\sl et al.}
attribute this flattening to star formation from cooling flow gas near
the galaxy centre.  This makes comparison of BCG \m2g data difficult,
unless one is careful to compare slopes outside emission-line regions
in all cases.
 
We can also compare our \m2g slopes with those of non-BCG ellipticals.
Couture \& Hardy (1988) found a mean \m2g gradient of $-$0.053 $\pm$
0.015 for six early-type galaxies; Davies, Sadler, \& Peletier (1993)
find a mean slope of $-$0.059 $\pm$ 0.022 for 13 normal ellipticals;
Gorgas {\sl et al.} (1990) obtain a mean slope of $-$0.058 $\pm$ 0.027
for 16 ellipticals; Davidge (1992) found a steeper gradient of
$-$0.081 $\pm$ 0.01 for 11 ellipticals.  Gonzalez \& Gorgas (1998; see
also Gonzalez \& Gorgas 1997) find a mean \m2g gradient of $-$0.055
$\pm$ 0.025 for 109 early-type galaxies.  Thus, our three BCGs have
slopes consistent with those found in non-BCG ellipticals.  
It seems likely that these BCGs have not experienced
significantly more mergers than normal ellipticals, as mergers are
expected to dilute abundance gradients (White 1980).

\setcounter{figure}{5}
\begin{figure*}
\epsfysize 8.5 truein
\hfil{\epsffile{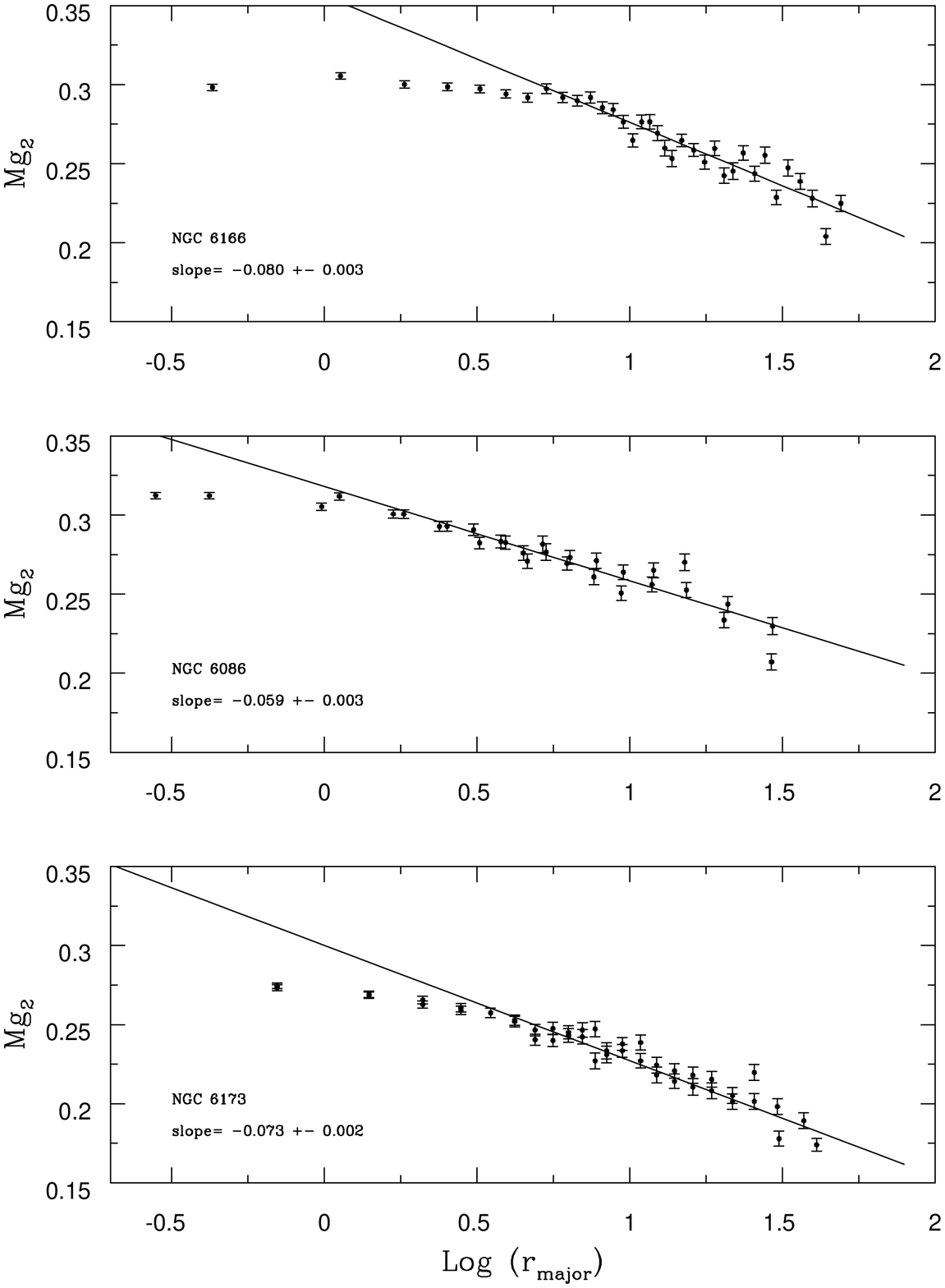}}\hfil
\caption{Mg$_2$ profiles for NGC 6166, NGC 6173, and NGC 6086.  
\m2g is plotted against the log of major axis radius.
For NGC
6173 and NGC 6086, data have been
folded about the major axis, while for NGC 6166
only the SW side of the major axis is shown because of problems with 
the secondary nucleus on the NE side.  The solid line in each plot
is a weighted least-squares
fit, with $|r| <$ 3 arcsec excluded for NGC 6173 and NGC 6086, and 
$|r| <$ 5 arcsec excluded for NGC 6166.}
\label{fig:mg2prof}
\end{figure*}

\section{Conclusions}

{\it Our main results are:}

\medskip

\noindent{\bf (1):~~}We find small rotation along the major axes of
NGC 6166, NGC 6173, and N6086, with V$_{rot}$ $\leq$ 50 km s$^{-1}$ at the
outermost observed major axis radii (30--40 arcsec) in the three
galaxies; the corresponding V/$\sigma$ ranges from 0.02--0.18.  This
small rotation is consistent with the nearly complete lack of rotation
found near the centres of a sample of 13 BCGs by Fisher {\sl et al.}
(1995).  This is consistent with merger models of BCG formation, but
data at larger radii are urgently needed, since numerical simulations
predict that mergers should contain significant amounts of angular
momentum beyond 2$-$3 r$_e$ (e.g. Hernquist 1993).

NGC 6173 and NGC 6086 have larger velocity gradients near the galaxy
centres, and NGC 6086 appears to have a counter-rotating core.  These
cores may arise from a previous merger/interaction (supported
by the shells seen in NGC 6173), or may be due to streaming in a
triaxial potential; further observations at other position angles will
be needed to decide between these two alternatives (Statler 1991).

\medskip

\noindent{\bf (2):~~~}NGC 6086 and NGC 6173 have slowly declining
velocity dispersion profiles.  In contrast, NGC 6166 has a rising
velocity dispersion: $\sigma$ increases from $\sim$ 325 km s$^{-1}$ at the
centre to $\sim$ 440 km s$^{-1}$ at 35\asec.

\medskip

\noindent{\bf (3):~~~}The Gauss-Hermite \h4 moment is positive and
roughly constant at 0.05 at all radii along the major axis of NGC 6173
and NGC 6086, and 0.1 along the major axis of NGC 6166.  Template
mismatch is unlikely to account for all of this offset, and we
conclude that there is possibly radial anisotropy present.  We will
confirm and quantify this suggestion with detailed modelling in a
later paper.

\medskip

\noindent{\bf (4):~~~} We find [OIII] and [NI] emission concentrated
within 5 arcseconds (3h$^{-1}$ kpc) of the centre of NGC 6166, with a
central velocity dispersion of $\sim$ 350 km s$^{-1}$, similar to the
central stellar velocity dispersion. The extent of the [OIII] emission
region is unresolved, but that of the [NI] may be marginally resolved.

\medskip

\noindent{\bf (5):~~}We find that [Mg/Fe] is larger in NGC 6166 and
NGC 6086 than in NGC 6173.  Although positive $h_4$ can lead to
systematic errors in [Mg/Fe], the magnitude of the offset appears too
large to be explained by such an error.  The difference in [Mg/Fe]
between galaxies with cD morphology and normal bright ellipticals, if
confirmed by further observations, would present a difficulty for
models in which the cDs were formed from mergers of normal galaxies of
any type.

\medskip

\noindent{\bf (6):~~} The \m2g gradients outside 3--5 arcsec range
from $-$0.06 to $-$0.08 for the three galaxies.  These slopes are
consistent with those found for BCGs without cooling flows or emission
lines by Cardiel {\sl et al.} (1997), and for non-BCG ellipticals.
More data are needed to definitively compare BCGs with non-BCG
ellipticals in this regard.

\medskip

We defer to a forthcoming paper a detailed dynamical modelling of
these three galaxies, but we note that (2) and (3) together imply that
the M/L ratio increases outwards in the NGC 6166 halo, and likely in
the other two galaxies as well.  Further support for a DM halo in NGC
6166 comes from X-ray data, e.g. Buote \& Canizares (1996).

\bigskip

Do these results favour one BCG 
formation model over others?  There are five main scenarios for
BCG formation: 
{\bf (i)}~ Deposition of cooling flow gas
from the ICM (e.g. Fabian 1994); 
{\bf (ii)}~Cannibalism of other
cluster galaxies (Ostriker \& Tremaine 1975);
{\bf (iii)~~} Tidal stripping of other
cluster galaxies, either by the cluster tidal field or by the BCG
itself (Richstone 1976; Moore {\sl et al.} 1996);
{\bf (iv)~~} Earlier origin in smaller
groups/subclusters,
which later merged to form the present-day cluster
(Merritt 1984, 1985; Tremaine 1990); and 
{\bf (v)~~} Dissipative collapse models (Larson 1975; Carlberg 1984a,b).
These scenarios are not 
necessarily mutually exclusive; 
for instance, many authors have suggested that 
mergers/cannibalism create the central body of the cD, while
tidal stripping accounts for their extended halos.

Dubinski (1998) has run an impressive simulation of cD formation,
using a large cosmological N-body simulation of cluster collapse with
hierarchical merging.  His `cluster' has a mass of 1.0 $\times$
10$^{14}$ M$_\odot$ and a line-of-sight velocity dispersion of 550
km s$^{-1}$, and he notes that it would be classified as a poor cluster or
large group by observers.  He finds that the final central giant
elliptical has a small rotation of $\sim$ 50 km s$^{-1}$, is radially
anisotropic in the outer regions, is aligned in position angle with
the `cluster' galaxy distribution, and has a slowly declining
major-axis velocity dispersion profile. These properties agree with
those observed in many BCGs, including our sample described in this
paper.  All three of our galaxies have little or no rotation, and have
significantly positive $h_4$ values which generally indicate radial
anisotropy.  NGC 6086 and NGC 6173 have velocity dispersion profiles
similar to Dubinski's remnant, and NGC 6173 and NGC 6166 are aligned
(to within 20 degrees) of their host cluster/subcluster galaxy
distributions and X-ray isophotes (Carter \& Metcalfe 1980; Dixon,
Godwin, \& Peach 1989; Muriel, Bohringer, \& Voges 1996; Buote \&
Canizares 1996).  However, more such simulations are needed to see if,
among other things, remnants with increasing dispersion profiles and
extended halos can be created.  Garijo, Athanassoula, \& Garcia-Gomez
(1997) have run a larger set of simulations of cD formation, with a
variety of initial conditions.  Some of the simulations produce
objects with extended halos and radial anisotropy, and all of the
final galaxies have declining velocity dispersion profiles.  Note,
however, that these simulations were run with all of the mass
initially bound to galaxies, probably not physically realistic.  The
radial anisotropy found by both authors is a result of the
`cosmological' mergers that are being modelled, where the initial
filamentary structure means that merging galaxies enter the central
object on mainly radial orbits.

Further constraints can be placed on formation models if one considers
the stellar abundance gradients. Our three galaxies all show modest
gradients, comparable with normal bright ellipticals (e.g. Davies,
Sadler, \& Peletier 1993). These gradients are not as steep as
predicted by simple dissipational collapse models (e.g. Larson 1975),
however they are steeper than we would expect from hierarchical
dissipationless merger models, which predict no gradients.  It is not
clear what the net effect of gaseous mergers will be, since on the one
hand mergers are expected to dilute existing abundance gradients
(White 1980), while on the other hand central star formation induced
by the merger could augment any gradient (see for instance Mihos \&
Hernquist 1994). In any case the gradients point to a formation
history not too different from that of normal bright ellipticals.

The [Mg/Fe] ratio in NGC 6166 and NGC 6086, and the difference between
these two BCGs and NGC 6173, provide important new clues as to the
differences between cDs and normal bright ellipticals. If this
difference is confirmed by further observations, this is a second
property, alongside the globular cluster specific frequency, in which
ellipticals differ from spirals and in which cDs are even more
extreme.  Fisher {\sl et al.} (1995) and Bender (1996) find that
[Mg/Fe] is correlated with velocity dispersion, so perhaps the large
[Mg/Fe] offsets seen in our BCGs are simply the continuation of this
correlation to the largest measured velocity dispersions.  The trend
of increasing [Mg/Fe] we see in our three BCGs (Figure 4) does agree
with their relative velocity dispersions.  A dissipative process is
most likely to give rise to these differences, although mergers as a
trigger for such a process cannot be ruled out; for instance Zepf and
Ashman (1993) interpret the bimodal colour (and hence metallicity)
distributions for globular clusters in some elliptical galaxies,
including the BCGs M87 and NGC 1399, as evidence for a second, merger
induced, epoch of cluster formation.

If we had data for NGC 6166 and NGC 6173 alone, it would be tempting
to draw a clear distinction between the two galaxies. NGC 6166 has a
rising velocity dispersion, and enhanced [Mg/Fe], a high globular
cluster specific frequency (Bridges {\sl et al.} 1996) and an X--ray
cooling flow. NGC 6173 has none of these things \footnote{The evidence
that NGC 6173 does not have a high globular cluster specific frequency
comes from a deep, unpublished, R band image taken at the same time as
the R-band image of Bridges {\sl et al.} (1996)}, yet is a galaxy with
a similar luminosity and surface brightness profile (Carter 1977). NGC
6086 appears to have a mixture of properties: it has a high [Mg/Fe],
yet does not have a rising dispersion profile. It is important to
establish which of these properties depend upon the environment of a
galaxy, and similar observations of galaxies with cD morphology in low
density environments (such as NGC 4839; Oemler 1976) are required.

It is not clear how to interpret the rising velocity dispersion
profile that we observe in NGC 6166.  One would expect a rising
dispersion profile under any model of BCG formation, since under
simple energy arguments the BCG dispersion profile should eventually
join that of the cluster as a whole (as is the case for NGC
1399/Fornax; see Section 1).  However, we note that the three BCGs
with increasing dispersions, and where X--ray data for the cluster
exist (NGC 6166, IC 1101, Abell 3266) are all in rich
clusters with smooth X-ray morphology.  There is no evidence for
substructure or peculiar BCG velocities in any of these clusters.  We
suggest that these are old, relaxed clusters in which the BCG is at
the centre of the cluster potential; these are the sorts of clusters
in which it will be most profitable to search for BCGs with increasing
velocity dispersions, in order to study the transition between the BCG
and the cluster. 

\section{ACKNOWLEDGMENTS}

We thank Hans-Walter Rix for the use of his program for absorption
line profile analysis, and for extensive advice on this subject.  We
would also like to thank Nicolas Cardiel, Javier Gorgas, Alfonso
Arag\'{o}n-Salamanca, Reynier Peletier, Jim Schombert and an anonymous
referee for comments
on various aspects of this paper.  Thanks also to Cardiel, Gorgas, and
Arag\'{o}n-Salamanca for sharing their data with us prior to
publication.  The Isaac Newton Telescope is operated on the island of
La Palma by the Royal Greenwich Observatory in the Spanish
Observatorio del roque de los Muchachos of the Instituto de
Astrofisica de Canarias.

\appendix

\section[INT montecarlo noise simulation]{Montecarlo simulations 
of the sensitivity of the recovered
kinematic parameters to noise}

\setcounter{figure}{0}
\begin{figure*}
\centerline{ \hbox{
\psfig{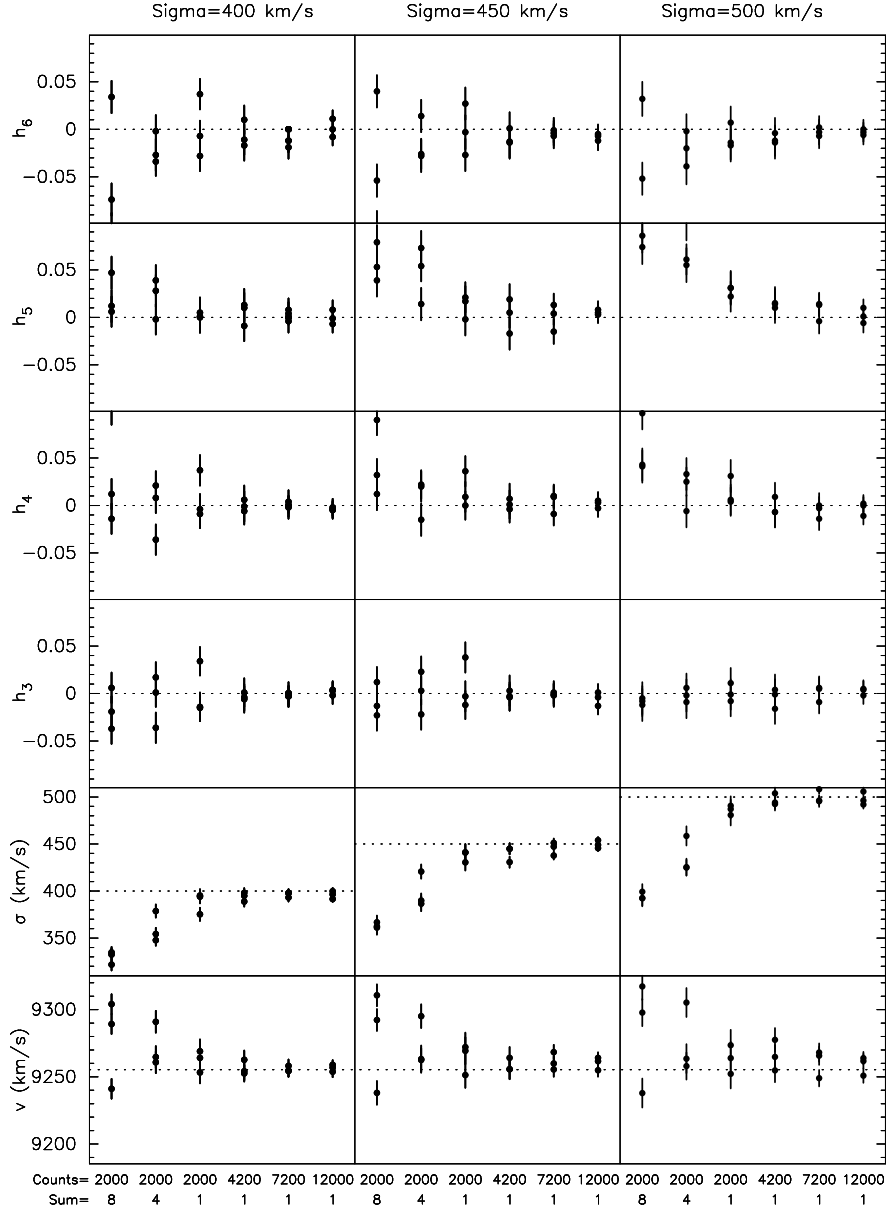}}}
\caption[The effect of noise on the recovery of the parameters for
NGC~6166]{The effect of noise on the recovery of the kinematic
parameters for NGC 6166. The
composite stellar template at the nucleus of NGC~6166 is convolved
with different model LOSVDs to make mock galaxy spectra.  Added to
these are Poisson noise and blank sky spectra. The kinematic
parameters are recovered from 3 mock spectra, each created with a
different noise seed and sections of blank sky. The panels from left
to right correspond to Gaussian LOSVDs with velocity dispersion
$\sigma$ of 400, 450 \& $500\,\kms$ respectively. In each panel,
results are shown for continuum levels of 2000, 4200, 7200 \& 12000
counts per pixel, labelled along the bottom with the number of rows of
sky summed. Up to 8 rows of sky spectra are added to the artificial
spectrum to mimmick the points farthest from the galaxy nucleus. The
model values are represented by the dotted lines, and the recovered
parameter by the solid dots.}
\label{fig:n6166monte}
\end{figure*}

\setcounter{figure}{1}
\begin{figure*}
\centerline{ \hbox{
\psfig{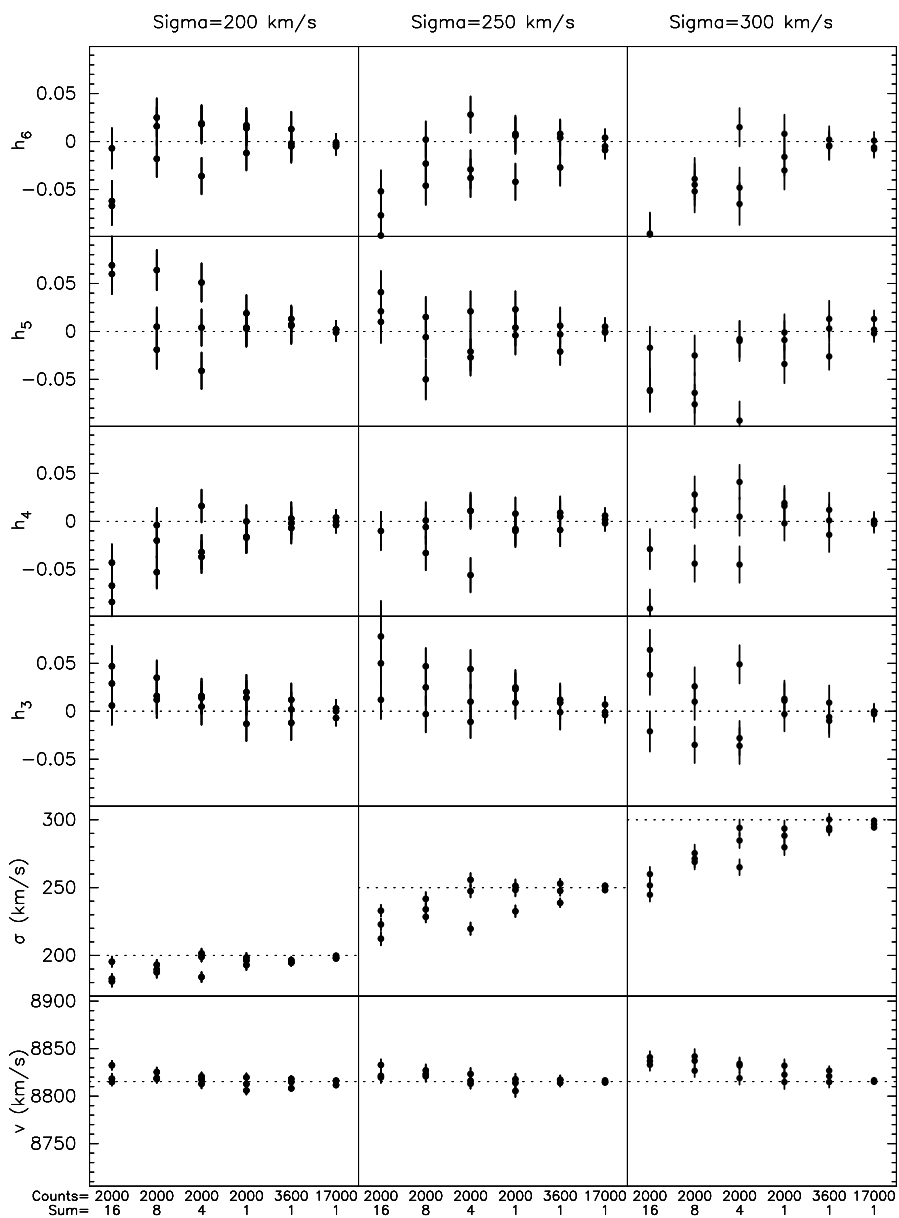}}}
\caption[The effect of noise on the recovery of the parameters for
NGC~6173.]{Same as Fig.~\ref{fig:n6166monte}, but for NGC~6173.  The
panels from left to right correspond to Gaussian LOSVDs with velocity
dispersion $\sigma$ of 200, 250 \& $300\,\kms$ respectively. In each
panel, results are shown for continuum levels of 2000, 3600 \& 17000
counts per pixel, labelled along the bottom with the number of rows of
sky summed. Up to 16 rows of sky spectra are added to the artificial
spectrum to mimmick the points farthest from the galaxy nucleus. The
model values are represented by the dotted lines, and the recovered
parameter by the solid dots.  }
\label{fig:n6173monte}
\end{figure*}

In this appendix the sensitivity of the kinematic parameters to noise
for NGC~6166, NGC~6173 \& NGC~6086 is investigated via Montecarlo
simulations of the effect of both Poisson noise and sky
residuals. Especially of interest is whether the lowered velocity
dispersions measured for points beyond $40\asec$ for NGC~6166 and
beyond $20\asec$ for NGC~6173 are significant. For this purpose
artificial spectra are generated by convolving the optimal stellar
templates at the centres of NGC~6166 \& NGC~6173 with perfect Gaussian
LOSVDs.

Poisson noise with different random seeds is added to the artificial
spectra.  Imperfect sky subtraction is investigated by adding blank
spectra extracted at large distances from the nucleus.  Up to 8 rows
are added to mimic the worst case of sky contamination for NGC~6166,
and 16 rows for NGC~6173. The kinematic parameters are recovered three 
times using {\tt kinematics}, each with different Poisson noise seeds
and sections of blank sky.

\subsection*{NGC~6166}

For NGC~6166, LOSVDs with $cz=9255.3\, \kms$ and $\sigma=400$, 450, \&
$500\,\kms$ are modelled.  The artificial spectra are scaled to
continuum levels of 12000, 7200, 4200 \& 2000 counts per pixel,
corresponding to distances of 0, 5, 8 \& $\sim 15\asec$ from the
nucleus respectively. Beyond $15\asec$ summation is required to
achieve 2000 counts per pixel. The number of spectra to be summed is
2, 4, 5 \& 8 for distances of approximately $22\asec$, $33\asec$,
$37\asec$ and $50\asec$ from the nucleus respectively.
The results of the experiments are plotted in
Fig.~\ref{fig:n6166monte}, where the left, centre and right panels
correspond to $\sigma=$ 400, 450 \& $500\,\kms$ respectively. For
$\sigma=400\,\kms$, the mean velocity $v$ can be determined to within
$2\,\kms$ at distances less than $5\asec$ from the nucleus. The
scatter increases to $6\,\kms$ at $r=15\asec$, $13\,\kms$ at
$r=33\asec$, and $27\,\kms$ at $r=50\asec$. In addition beyond
$33\asec$ there is a small systematic offset of $\sim 20\,\kms$
towards higher velocity. For broader LOSVDs, the behavior is similar,
except that the recovering accuracy is lower.

If the velocity dispersion is $\sim 400\,\kms$, it can be determined
to within $\sim 3\,\kms$ at distances less than $5\asec$ from the
nucleus.  Its scatter increases with distance. In addition, as the
contribution of the sky increases with decreasing $S/N$, the recovered
value is systematically lower.  The offset is about $-12\,\kms$ at
$r=15\asec$, $-40\,\kms$ at $r=33\asec$, and $-70\,\kms$ at
$r=50\asec$. The behavior is similar if the LOSVD is broader, except
that the magnitude of the systematic offset is larger. At the extreme,
the recovered value for a Gaussian LOSVD with $\sigma=500\,\kms$ at
$r=50\asec$ is $100\,\kms$ lower.

In general, the Gauss-Hermite moments \h3 and \h4 can be recovered
reasonably well if $\sigma = 400\,\kms$, to within $\sim 0.025$ at
distances up to $33\asec$.  Their scatter increases with
$\sigma$. No obvious systematics are introduced in  \h3 even at the
outermost points, or when $\sigma$ is increased to $500\,\kms$.  A
small positive offset of order $0.03$ is introduced in  \h4 for
data points beyond $33\asec$, and this tends to increase with
$\sigma$.  The Gauss-Hermite moments \h5 and \h6 can be recovered well
to within 0.01 in the inner $15\asec$, and to $\sim 0.02$ at $r=33
\asec$. Again, the scatter increases with $\sigma$. In addition a
significantly positive offset of $\sim +0.05$ is introduced to the $\h5$
at the outer points if sigma is increased beyond $400\,\kms$.

The conclusions from this exercise are that at distances beyond $\sim
33\asec$, where summation of more than 4 spectra are required,
systematics start to creep in as the sky contribution increases. At such
distances, the recovered velocity dispersions are systematically
lower. This shows that we are probably unable to measure LOSVDs
broader than $\sim 400\,\kms$ beyond these distances, and the low
($\sim 320 \kms$) velocity dispersions beyond $40\asec$ actually
measured for NGC~6166 are entirely consistent with expectation. For
these reasons, we measure $v$ up to $50\asec$, and $\sigma$, \h3,
\h4, h5 \& h6 up to $36\asec$ from the nucleus.

\subsection*{NGC~6173}

For NGC~6173, LOSVDs with $cz=8815.3\,\kms$ and $\sigma=200$, 250, \&
$300\,\kms$ are modelled. The artificial spectra are scaled to
continuum levels of 17000, 3600 \& 2000 counts per pixel,
corresponding to distances of 0, 6, \& $\sim 9\asec$ from the nucleus
respectively. Beyond $9\asec$ summation is required to achieve 2000
counts per pixel. The number of spectra to be summed is 4, 8 \& 16
for distances of approximately $19\asec$, $30\asec$ and
$46\asec$ from the nucleus respectively.
The results of the experiments are plotted in
Fig.~\ref{fig:n6173monte}, where the left, centre and right panels
correspond to $\sigma=$200, 250 \& $300\,\kms$ respectively. For
$\sigma=200\,\kms$, the mean velocity $v$ can be determined to within
$4\,\kms$ at distances less than $6\asec$ from the nucleus. The
scatter increases to $6\,\kms$ at $r=9\asec$ and $8\,\kms$ at
$r=46\asec$. In addition beyond $30\asec$ there is a small systematic
offset of $\sim +6\,\kms$. For broader LOSVDs, the behavior is
similar, except that the magnitude of the scatter and the systematic
offset is slightly higher.

If the velocity dispersion is $\sim 250\,\kms$, it can be determined
to within $\sim 6\,\kms$ at distances less than $6\asec$ from the
nucleus. Its scatter increases with distance. In addition, as the
contribution of the sky increases with decreasing $S/N$, the recovered
value is systematically lower.  The offset is about $-9\,\kms$ at
$r=19\asec$, $-15\,\kms$ at $r=30\asec$, and $-27\,\kms$ at
$r=46\asec$. The magnitude of the systematic offset is larger if the
LOSVD is broader, and smaller if the LOSVD is narrower. At the
extreme, the recovered value for a Gaussian LOSVD with
$\sigma=300\,\kms$ at $r=46\asec$ is $48\,\kms$ lower.

In general, the Gauss-Hermite moments \h3 and \h4 can be recovered
reasonably well if the $S/N$ is high. If $\sigma=250\,\kms$, they can
be determined to within $\sim 0.02$ at distances up to $9\asec$.
Their scatter increases with $\sigma$ and sky contribution. A small
positive offset in  \h3 is introduced with increasing sky
contribution.  No significant systematics are introduced in \h4, except
for the outermost points at $r \sim 46\asec$, where \h4 tends to be
lower. The Gauss-Hermite moments \h5 and \h6 can be recovered well
only near the galaxy centre ($r< 9 \asec$). The scatter increases with
$\sigma$ and increasing sky contribution. If $\sigma = 250\,\kms$,
there are no obvious systematics in \h5 with increasing sky
contribution. However, at distances larger than $19\asec$  \h5 is
scattered upwards if $\sigma=200\,\kms$, and downwards if
$\sigma=300\,\kms$. \h6 tends to be scattered downwards  at
distances larger than $19\asec$ for all $\sigma$.

The tests for NGC~6173 should also be applicable to NGC~6086, as both
galaxies have similar velocity dispersions, and are at similar
distances.  For NGC~6086, the continuum levels are 14600 and 2000
counts per pixel at $r=0\asec$ and $5\asec$. Beyond $5\asec$ summation
is required. The number of spectra to be summed is 4, 6 \& 15 for
distances of approximately $12\asec$, $15\asec$ and $30\asec$ from the
nucleus respectively.
 
The conclusions from this exercise are that at distances beyond $\sim
19\asec$, where summation of more than 4 spectra are required,
systematics start to creep in as the sky contribution increases. At such
distances, the recovered velocity dispersions are systematically
lower, but the problem is not as severe as in NGC~6166. This shows
that the fall in $\sigma$ beyond $20\asec$ is probably more gradual in
reality. The significance of the Monte Carlo tests is that the $\sigma$
in NGC~6173 is consistent with being flat or falling beyond $20\asec$,
but inconsistent with rising outwards. We measure $v$, $\sigma$ \&
\h3 up to $40\asec$, \h4 up to $25\asec$, and \h5 \& \h6 up to $19\asec$
from the nucleus.

\subsection*{Recovery of High Values of $h_4$}

\begin{figure*}
\centerline{ \hbox{
\psfig{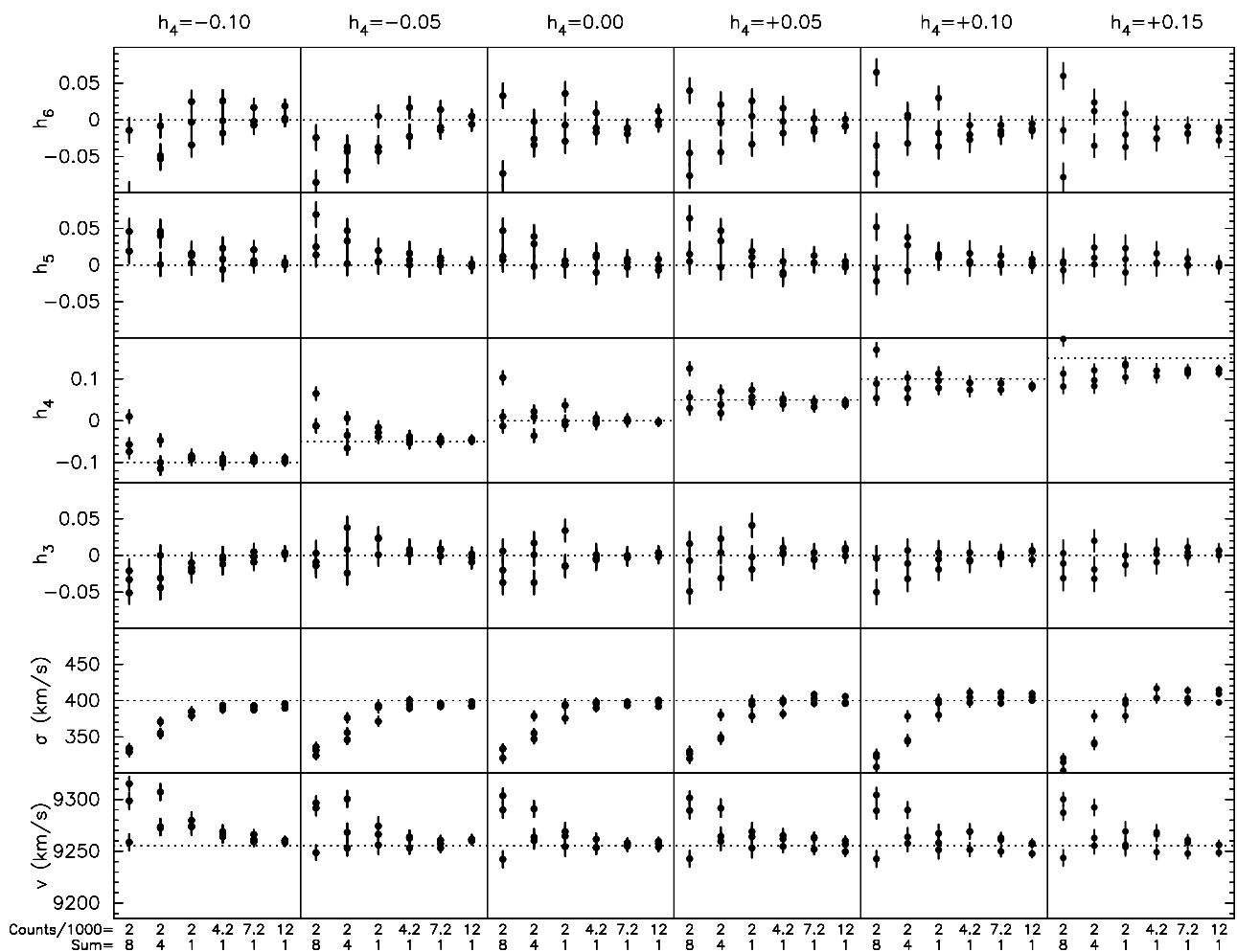}}}
\caption[The effect of noise on the recovery of $h_4$ for NGC
6166]{Same as Fig.~\ref{fig:n6166monte}, but the six columns now
represent different input values of $h_4$, each for $\sigma=400\,\kms$.  
The
panels from left to right correspond to LOSVDs with $h_4$
of -0.10, -0.05, 0.00, +0.05, +0.10 and +0.15 respectively. All other 
parameters are as for the left hand panel of Fig.~\ref{fig:n6166monte}.
}
\label{fig:h4test}
\end{figure*}

The tests described above show that the systematic errors on the recovery
of $h_4$ caused by noise are small in the case of $h_4$ = 0. It is important
also to ask what other values of $h_4$ can be recovered by our techniques.
For NGC 6166 we model LOSVDs with $cz=9255.3\,\kms$ ans $\sigma=400\,\kms$,
and $h_4=$-0.10, -0.05, 0.00, +0.05, +0.10 and +0.15 respectively, and the 
results are plotted in Fig.~\ref{fig:h4test}. The systematic bias on $\sigma$
at low $S/N$ is marginally worse for positive $h_4$ (second panel up in each 
column), and the main result of this exercise is that there is a systematic
underestimation of $|h_4|$ for $|h_4| \ge 0.1$. Our measurement of high
$h_4$ in NGC 6166 could possibly still be a systematic underestimate.

\subsection*{The Effect of Template Choice}

\begin{figure*}
\centerline{ \hbox{
\psfig{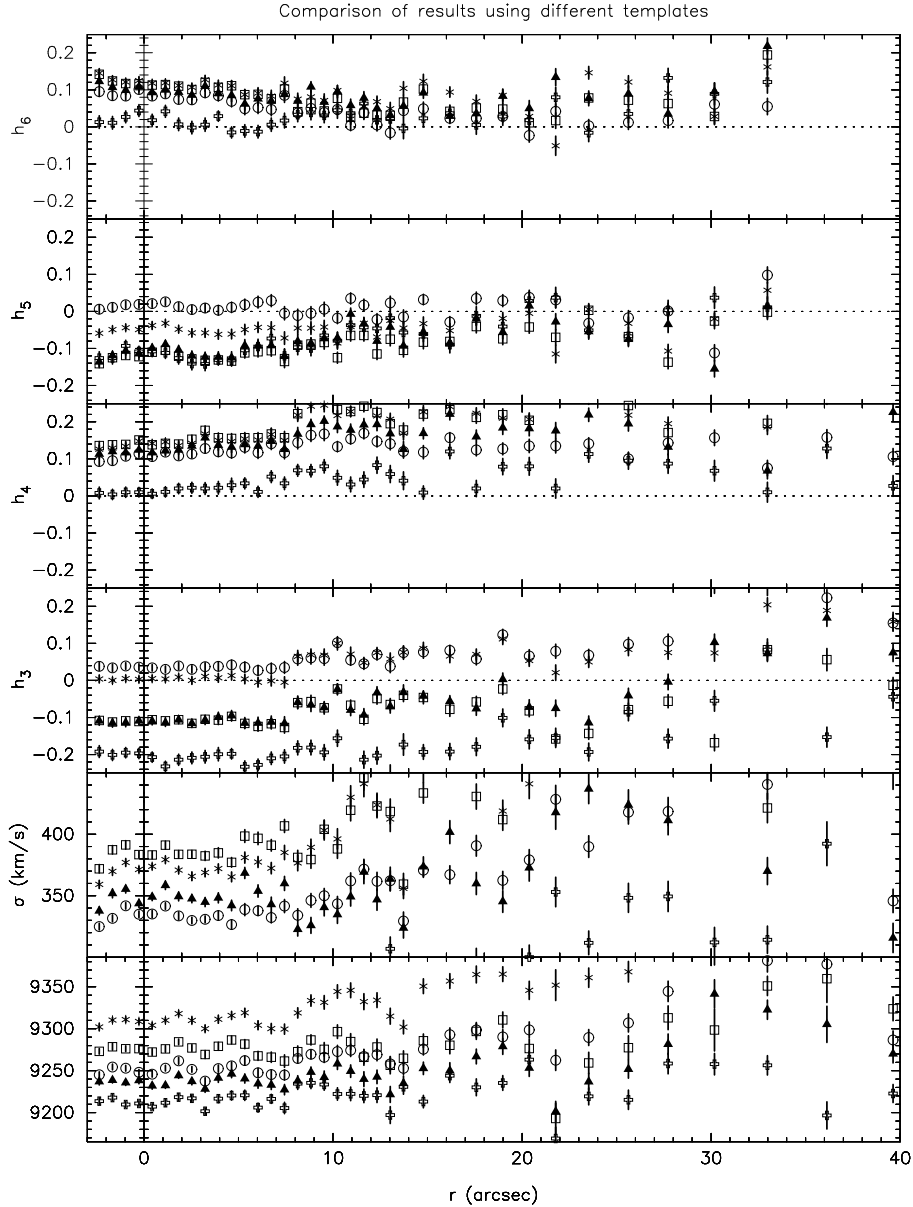}}}
\caption[The effect template choice on the derived kinematic
parameters]{Kinematic parameters for NGC 6166 derived using individual
stellar template spectra: HR7555 
(G6III, open squares); HR4864 (G7V, filled triangles); HR6674
(K0III, crosses); HR4676 (K4III, open circles) and HR8057 (M1III, thin 
rectangles).}
\label{fig:templatetest}
\end{figure*}

Our kinematic parameters are all determined using optimal best fit templates
constructed from 28 stellar spectra. To demonstrate that this procedure does 
not itself introduce systematic effects into the kinematic parameters, here
we repeat the analysis for NGC 6166 using individual template stars. Moreover
to demonstrate the effect of template choice on the derived parameters we
have carried out this exercise with templates with a wide range of spectral 
type. In Fig.~\ref{fig:templatetest} we plot the kinematic parameters for
NGC 6166 derived from the five individual templates: HR7555 
(G6III, open squares); HR4864 (G7V, filled triangles); HR6674
(K0III, crosses); HR4676 (K4III, open circles) and HR8057 (M1III, thin 
rectangles). The trends in the derived parameters are identical to those in
Fig.~\ref{fig:hermplot}, although there are differences in the zero points 
which depend upon the template choice. The difference in the velocity zero 
points is not real, it is because this procedure determines the velocity 
relative to the first star in the list of templates, if there is only one
it depends upon the radial velocity of that star. For $\sigma$ and all of the 
Hermite terms the M1III star HR8057 gives a very different zero point,
the dispersions are so much lower that many of them are off the bottom of the 
dispersion panel in Fig.~\ref{fig:templatetest}. However this star is a very
poor fit to the galaxy spectra and does not contribute much to the optimal
composite templates. For the other templates, $h_4$ is tightly defined around
+0.10, suggesting that template mismatch errors do not contribute to our
large positive values of $h_4$. On the other hand there are large zero point 
differences in $h_3$ between the G and K star templates, to get reliable values
of $h_3$ it is important that the templates used fit well. 

The velocity dispersion zero points differ with a total spread of 50 $\kms$
(again neglecting HR8057),
so the uncertainty of the zero point of the velocity dispersion profile due to
template mismatch might be estimated at 25 $\kms$. 

\section[Lick absorption line indices]
{Procedure for measuring Lick absorption line
indices}

In this appendix we describe the procedures for obtaining absorption
line indices in the Lick system.  Line indices are
extracted from individual, de-redshifted spectra following the
procedures outlined in Worthey \& Ottaviani (1997), adopting the
passband definitions of Worthey {\sl et al.} (1994). The blue-ward
continuum bandpass for \m2g is redefined to be
4931.625$-$4957.625 \ang~ instead of 4895.125$-$4957.625 \ang,
because the spectra of the template stars only start from about
4931 \ang.  The difference between the Mg index 
\m2g$^{alt}$ under this alternative definition, and the Lick \m2g index
is tiny (see below). The molecular
band  \m2g$^{alt}$ is measured in pseudo magnitudes,
whilst the atomic absorption line indices  \fe52 and \f53e 
are measured as equivalent widths in Angstroms.
 
\begin{table*}
\footnotesize
\centering
\caption{The bandpass wavelengths of absorption line indices 
\m2g, \fe52, and \f53e defined by Worthey \etal
(1994), with our alternative definition for 
\m2g$^{alt}$. The latter is measured and transformed to \m2g
via the scaling relation described in the text. }
\label{tab:n6166lickdefs}
\vspace{0.5cm}
\begin{tabular}{lccc}
\hline
Feature & Type & Central Bandpass  & Continuum Bandpasses  \\
        &      &      (\ang)              &       (\ang) \\
\\
\m2g$^{alt}$&Molecular band       & 5154.125--5196.625  & 4931.625--4957.625,
5301.125--5366.125 \\
\m2g      &Molecular band         & 5154.125--5196.625  & 4895.125--4957.625,
5301.125--5366.125 \\
\fe52 &Atomic absorption line & 5245.650--5285.650  & 5233.150--5248.150,
5285.650--5318.150 \\
\f53e &Atomic absorption line & 5312.125--5352.125  & 5304.625--5315.875,
5353.375--5363.375 \\
\hline
\end{tabular}
\normalsize
\end{table*}
 
First, our spectra are broadened to a resolution of $8.5 \ang$~ FWHM, in
common with the Lick/IDS resolution in this part of the spectrum. In
NGC 6166, there is significant [NI] emission at $-8.7\asec < r <
8.1\asec$. As the [NI] emission line is halfway inside the \m2g
passband, its flux can artificially lower the \m2g by $7\%$ at the
nucleus.  Therefore the flux contribution between rest wavelengths
5183.125 $\ang$ and 5196.526 $\ang$ is measured from the residuals of the
kinematic fitting and subtracted off the flux in the \m2g passband.
 
The indices are then corrected to a zero dispersion system by applying
correction factors. These are estimated by broadening the optimal
stellar template of NGC~6166 to different velocity dispersions, and
are summarized in table~\ref{tab:n6166dispcor}.  The correction for
\m2g is insignificant even at large dispersions, whilst \f53e
requires a rather large correction. These corrections are not sensitive
to the choice of template used to determine these corrections, in the worst 
case (\f53e at high velocity dispersion) the total range in the factor
determined using a range of templates is 0.034, implying an error in the factor
as determined from the best fit template of at most 0.01.
 
\begin{table}
\footnotesize
\centering
\caption{Velocity dispersion correction factors for different line
indices. The measured indices should be divided by these to transform
back to the zero dispersion system.}
\label{tab:n6166dispcor}
\vspace{0.5cm}
\begin{tabular}{cccc}
\hline
$\sigma$ & \m2g & \fe52 & \f53e \\
 ($\kms$) & & & \\
\\

160 &  0.998 & 0.938 & 0.895\\
180 &  0.998 & 0.924 & 0.868\\
200 &  0.997 & 0.909 & 0.842\\
220 &  0.997 & 0.895 & 0.815\\
240 &  0.997 & 0.879 & 0.785\\
260 &  0.996 & 0.864 & 0.756\\
280 &  0.996 & 0.849 & 0.727\\
300 &  0.995 & 0.834 & 0.697\\
320 &  0.995 & 0.819 & 0.668\\
340 &  0.994 & 0.804 & 0.640\\
360 &  0.993 & 0.788 & 0.612\\
380 &  0.991 & 0.773 & 0.586\\
400 &  0.990 & 0.758 & 0.560\\
420 &  0.988 & 0.742 & 0.536\\
440 &  0.987 & 0.726 & 0.512\\

\hline
\end{tabular}
\normalsize
\end{table}
 
\subsection
[Conversion to Lick system]{Conversion from alternative system to Lick 
system}

\begin{figure}
\epsfxsize 6.0 truecm
\hfil\rotate[r]
{\epsffile{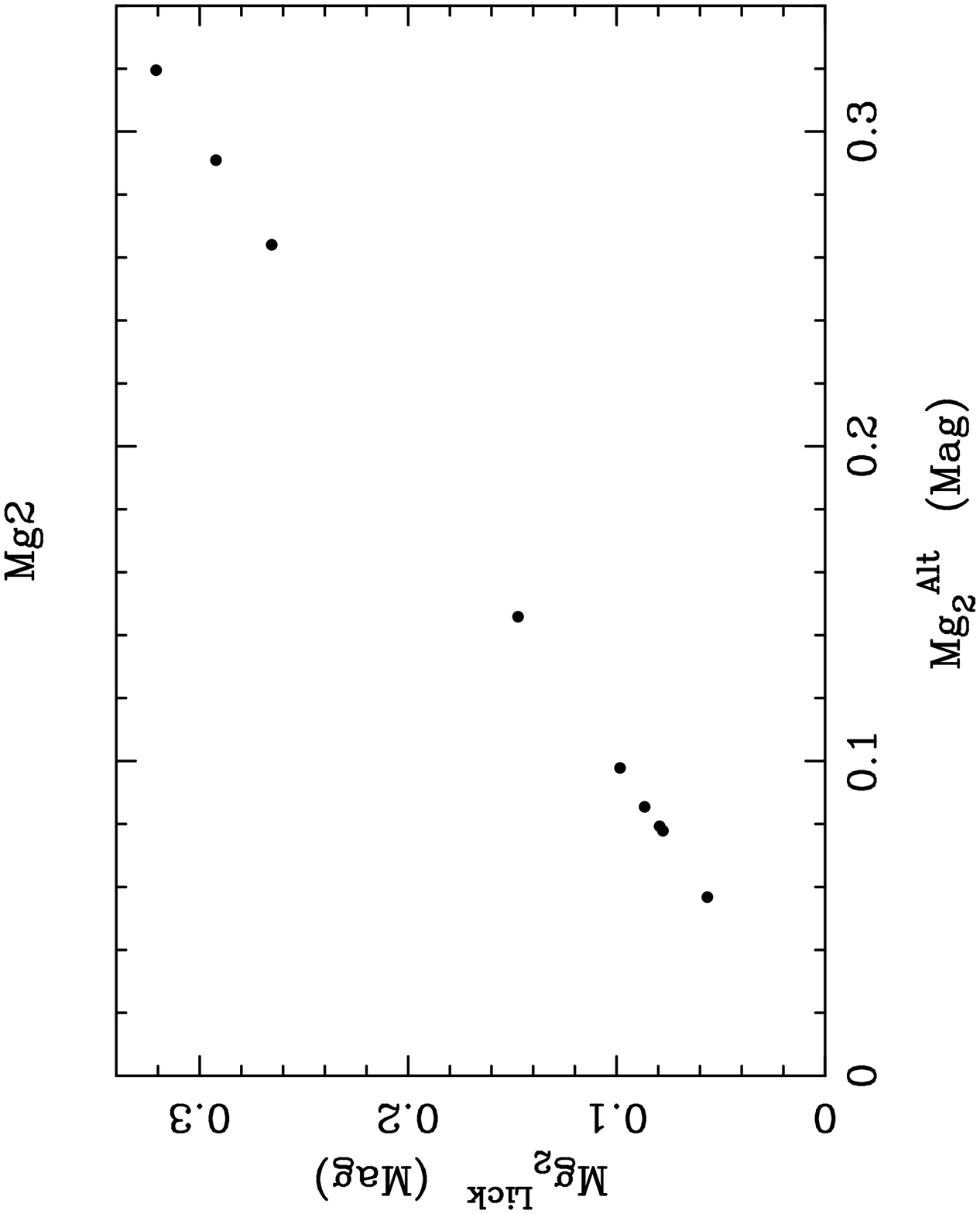}}\hfil
\caption[Lick Mg indices vs alternative Mg]{
\m2g measured according to the Lick definition is
plotted against that measured according to our alternative
definition.}
\label{fig:mglickvsalt}
\end{figure}

To convert \m2g$^{alt}$ to \m2g, a scaling
relationship is required.  This is obtained by comparing the \m2g 
indices for 9 template stars measured under the Lick and our
alternative definition. The template stars were observed on the AAT in
March 1996 using the RGO spectrograph, and their spectra span the
wavelength range 4850 -- 5610 $\ang$, and are broadened to $8.5 \ang$
FWHM resolution. The fact that these spectra are taken with a
completely different instrument and telescope does not pose a problem,
as the only variable is the change in bandpass definition. The \m2g 
indices under Worthey's definition are plotted against those under the
alternative definition in Fig.~\ref{fig:mglickvsalt}, and the
following relationship is obtained by a linear least-squares fit:
\begin{eqnarray}
Mg_2 &=& 1.005 Mg_2^{alt}
\end{eqnarray}
Because the changes introduced by redefining the continuum bandpass
are only 0.5\% for \m2g, they are 
unlikely to be a major source of error.

\subsection[Zero points]{Lick zero points}
 
Because our spectra are not flux calibrated, there might be a zero point
difference between our measured indices and the Lick values, due to the
curvature of the response function. Of our standard stars, only one, HR 5277
has been measured on the Lick system. Worthey (private communication) ascribes
low weight to the observations of this particular star, and in any case using
observations of a standard star to determine the zero points for galaxies is
subject to systematic error, due to the substantial redshifts of our
galaxies, and the corresponding difference in the grating settings for the
observations. Thus we are unable to tie our measurements to the Lick system
directly. On the other hand Cardiel {\sl et al.} (1997) have measurements
which we, and they, believe to be on the Lick system. Figure
~\ref{fig:uscardiel} demonstrates good agreement between the \m2g values
in the region in which the data of Cardiel {\sl et al.} are reliable, and
thus we believe that the zero point offset between our values and the Lick
system is small, at least in \m2g. 

Trager {\sl et al.} (1998) publish Lick indices for a 1.4 x 4 arcsecond
aperture centred on the nucleii of NGC 6086 and NGC 6166.
For NGC 6086 they find $Mg_2 = 0.309 \pm 0.008$; $Fe_{5270} = 3.32 \pm 0.32 
\AA$; $Fe_{5335} = 2.71 \pm 0.47 \AA$. For NGC 6166 they find $Mg_2 = 0.315 
\pm 0.010$; $Fe_{5270} = 3.21 \pm 0.34 \AA$; $Fe_{5335} = 2.01 \pm 0.47 \AA$. 
These are consistent with our results, although the error bars on their iron 
line equivalent widths are disappointingly large. Although some uncertainty
remains as to the zero point of the iron line indices, it seems that our
indices are within 0.01 magnitudes of the Lick system in the case of \m2g,
and within 0.3 Angstrom in the case of the combined iron line index.

\end{document}